\documentclass[11pt]{article}
\usepackage{graphicx}
\usepackage{amsfonts}
\usepackage{amsmath}
\usepackage{amssymb}
\usepackage{url}
\usepackage{fancyhdr}
\usepackage{indentfirst}
\usepackage{enumerate}
\usepackage{multirow}
\usepackage{booktabs}
\usepackage{multirow}
\usepackage{amsthm}
\usepackage{color}
\usepackage{natbib}
\usepackage{dsfont}
\usepackage[colorlinks=true,citecolor=blue]{hyperref}
\usepackage{caption}      
\usepackage{subcaption}   
\usepackage[ruled,vlined]{algorithm2e}

\usepackage[misc]{ifsym}
\def\d{\,\mathrm{d}}
\def\laweq{\buildrel \d \over =}

\newcommand{\VaR}{\mathrm{VaR}}

\newcommand{\TVaR}{\mathrm{TVaR}}

\newcommand{\E}{\mathbb{E}}
\newcommand{\R}{\mathbb{R}}

\newcommand{\p}{\mathbb{P}}

\newcommand{\id}{\mathds{1}}

\newcommand{\cI}{\mathcal{I}}

\newcommand{\esssup}{\mathrm{ess\mbox{-}sup}}
\newcommand{\essinf}{\mathrm{ess\mbox{-}inf}}
\renewcommand{\ge}{\geqslant}

\renewcommand{\geq}{\geqslant}
\renewcommand{\leq}{\leqslant}
\renewcommand{\epsilon}{\varepsilon}
\theoremstyle{plain}
\newtheorem{theorem}{Theorem}[section]

\newtheorem{lemma}{Lemma}[section]
\newtheorem{proposition}[theorem]{Proposition}
\theoremstyle{definition}
\newtheorem{definition}{Definition}
\newtheorem{assumption}{Assumption}
\newtheorem{example}{Example}[section]

\theoremstyle{remark}
\newtheorem{remark}{Remark}
\theoremstyle{definition}

\renewcommand{\cite}{\citet}

\begin{document}
\title{Performance-based variable premium scheme and reinsurance design }

\author{
Ziyue Shi\thanks{Department of Statistics and Actuarial Science, University of Waterloo, Canada. \Letter~\texttt{ziyue.shi@uwaterloo.ca}.}\and
David Landriault,\thanks{Department of Statistics and Actuarial Science, University of Waterloo, Canada. \Letter~\texttt{david.landriault@uwaterloo.ca}.}\and 
Fangda Liu,\thanks{Department of Statistics and Actuarial Science, University of Waterloo, Canada. \Letter~\texttt{fangda.liu@uwaterloo.ca}.} 
}

\date{\today
\thanks{\baselineskip=0.75\normalbaselineskip   D.~Landriault and F.~Liu acknowledge financial support from the Natural Sciences and Engineering Research Council of Canada (RGPIN-04011, RGPIN-2020-04717, DGECR-2020-00340).} }
\maketitle

\begin{abstract}
In the literature, insurance and reinsurance pricing is typically determined by a premium principle, characterized by a risk measure that reflects the policy seller's risk attitude.
Building on the work of \cite{meyers1980analysis} and \cite{CCT2016}, we propose a new performance-based variable premium scheme for reinsurance policies, where the premium depends on both the distribution of the ceded loss and the actual realized loss. Under this scheme, the insurer and the reinsurer face a random premium at the beginning of the policy period.
Based on the realized loss, the premium is adjusted into either a ``reward'' or ``penalty'' scenario, resulting in a discount or surcharge at the end of the policy period. 
We characterize the optimal reinsurance policy from the insurer's perspective under this new variable premium scheme. 
In addition, we formulate a Bowley optimization problem between the insurer and the monopoly reinsurer. Numerical examples demonstrate that, compared to the expected-value premium principle, the reinsurer prefers the variable premium scheme as it reduces the reinsurer's total risk exposure. 

\noindent \begin{bfseries}Key-words\end{bfseries}:
Variable premium scheme, reinsurance pricing, distortion risk measure,  optimal reinsurance policy, Bowley optimal
\end{abstract}

\section{Introduction}

Reinsurance is a crucial risk management tool that allows insurance companies to mitigate risk and expand their operations. In exchange for transferring part of their risk, insurers pay a reinsurance premium to the reinsurer, which typically increases with the size of the ceded loss. The risk of a reinsurance contract consists of both the retained loss and the premium, therefore, there is a trade-off between the amount retained and the premium, which stimulates the research for optimal reinsurance problems. The foundational work in the theoretical study of optimal reinsurance was led by \cite{borch1960attempt}, which seeks the optimal policy to minimize the variance of the insurer's risk. 
Since Borch's pioneering model, the optimal reinsurance framework has been extended in various directions, including discussions on the admissible policy set, the optimization objective, and the premium principle.
The admissible pricing rules include, for example, linear pricing criteria as in \cite{borch1962equilibrium} and \cite{buhlmann1984general}, mean-variance pricing criteria as in \cite{Kaluszke2001} and \cite{cai2008optimal}, Choquet pricing criteria as in \cite{Young1999} and \cite{de2003choquet}, and convex premium principle as in \cite{deprez1985convex} and \cite{ghossoub2023bowley}. In nearly all studies on optimal reinsurance, premiums are calculated based on a risk measure, which depends on the distribution of the ceded loss to the reinsurer. 

In insurance practice, certain pricing methods, such as retrospective rating, consider both the distribution of the ceded loss and the actual realized loss. Retrospective rating adjusts the premium based on the policyholder's actual losses during the policy period, with higher premiums reflecting greater losses. This approach incentivizes policyholders to manage risks effectively. It is commonly applied in workers' compensation, general liability, and auto liability insurance. Under a retrospective rating plan, the premium increases linearly with the realized loss, but to prevent the premium from being excessively low for the seller or prohibitively high for the buyer, the plan includes both a minimum threshold and a maximum cap. Determining parameters in this premium can be complex and the premium introduces additional randomness beyond the retained and ceded losses for both the insurer and reinsurer.
Despite its practical application in insurance, the retrospective rating plan premium is seldom addressed in theoretical research due to its complexity. 
\cite{mahler1976interaction} examines the miscalculation in the standard retrospective rating premiums due to the failure to account for the cross effects of various elements when calculating the basic premium.
\cite{meyers1980analysis} proposes a modified formula for calculating the basic premium that considers the claim severity distribution. \cite{lee1988mathematics} illustrates the mathematical idea behind the retrospective rating by a graphical approach. 
In \cite{CCT2016}, the authors investigate an optimal retrospective rating plan by minimizing the policyholder's risk exposure in the sense of convex order. They approach the problem from the policyholder's perspective and conclude that the insurer favors a policy priced using the expected value premium principle, a specific form of the retrospective rating plan premium.
We highlight that, under the retrospective rating plan premium, the premium is random at the time of signing the policy. The actual amount can only be observed at the end of the policy period when the realized loss amount is available.  The randomness of the retrospective rating plan premium is the key feature distinguishing it from other classical premium principles which determine the premium as a constant for an insurance policy.

In the presenting work, we introduce a new performance-based premium scheme for reinsurance policies, referred to as the ``reward-and-penalty" variable premium scheme. Inspired by the retrospective rating plan premium, the proposed scheme depends on the realized loss, making it a random variable for both the insurer and reinsurer at the beginning of the reinsurance policy period. 
As the policy seller,the reinsurer establishes the premium structure at the outset, such that the insurer pays a lower premium if the realized loss is small and conversely, a higher premium if a significant loss occurs.
In other words, the actual premium paid by the insurer is contingent on their performance during the policy period. 
The proposed variable premium scheme may motivate the insurer to strategically choose the reinsurance policy to avoid a large amount of premium. Furthermore, the insurer's cautions on choosing a reinsurance policy can also benefit the reinsurer through reducing the reinsurer's risk exposure.
Therefore, beyond analyzing the insurer's response to this premium structure, it is equally valuable to investigate the reinsurer's behavior under a retrospective rating plan.
In summary, this work makes three key contributions. First, we propose a new variable premium scheme based on the performance of the insurable loss. Second, we examine the insurer's optimal strategy when applying this scheme. The introduction of additional randomness from the premium complicates the insurer's optimization problem, making it more mathematically challenging. Lastly, we analyze the impact of this variable premium scheme on the reinsurer's risk exposure.

It is worth noting that the insurer and the reinsurer often have conflicting interests in a reinsurance policy. One effective approach to addressing these conflicts is by investigating the Bowley optimum between the two parties. The Bowley optimum follows a two-step process. Initially, the policy buyer chooses an optimal indemnity function in accordance with the seller's pricing rule to maximize their objective. Subsequently, the monopoly seller responds by selecting optimal pricing criteria to maximize their own objective, taking into account the anticipated reactions of the buyer. The Bowley optimum was first introduced into reinsurance by \cite{buhlmann1968marktverhalten}, who derive the solution for one insurer and one reinsurer under quadratic utility assumptions. This work was later extended by \cite{gerber1984chains} and \cite{chan1985reinsurer}.
For decades, the Bowley solution remained unexplored until \cite{cheung2019risk} extend the foundational work by incorporating a general pricing rule and a comprehensive risk management tool. More recently, \cite{chi2020bowley} examine a scenario where the insurer is constrained by a two-moment condition set by the reinsurer, while \cite{li2021bowley} apply the Bowley solution to a mean-variance problem. In addition, \cite{boonen2021bowley} and \cite{boonen2022bowley} revisit the Bowley problem with asymmetric information on the insurer's distortion risk measure. \cite{ghossoub2023bowley} explore a subclass of convex premium principles, and \cite{boonen2023bowley} investigate the relationship between Bowley optimality and Pareto efficiency in optimal reinsurance problems.
In this work, we propose a Bowley reinsurance problem incorporating a variable premium scheme. Through numerical analysis, we demonstrate that this variable premium can reduce the reinsurer's total risk exposure compared to the expected-value premium principle, which is a constant premium scheme.

The following part is organized as follows. 
In Section \ref{proj2-sec1}, we introduce the reward-and-penalty variable premium scheme and discuss its properties.
Section \ref{sec:insurer's opt prob} examines the insurer's optimization problem under the proposed scheme, providing a characterization of the optimal reinsurance strategy when the insurer uses either a general risk measure or a distortion risk measure to quantify total risk exposure.
In Section \ref{proj2-sec3}, we establish a Bowley optimization problem between the insurer and the reinsurer, and numerically analyze the impact of the variable premium scheme from the reinsurer's perspective. Finally, Section \ref{sec:conclusion} presents concluding remarks.

\section{Reward-and-penalty variable premium scheme}\label{proj2-sec1}

 \subsection{Notation}

We work with an atomless probability space $(\Omega, \mathcal F, \p)$.
Let $L^q$ be the set of all random variables in $(\Omega, \mathcal F, \p)$ with finite $q$-th moment, $q\in (0,\infty)$.
We assume that an insurable risk $X$ has a finite mean and should be a pure loss, that is, $X \in L^1$ and $X \geq 0 $ a.s.
We use $F_X$ and $S_X = 1 - F_X $  to represent the distribution function and the survival function of $X$, respectively. The left-continuous inverse of $F_X$ is given by $$F_X^{-1}(p)=\inf\{x\in \R: F_X(x)\ge p\},~p\in (0,1).$$ 
The mappings $\essinf(\cdot)$ and $\esssup(\cdot)$ on $L^1$ represent the essential infimum and the essential supremum of a random variable, respectively.
The random variables $X$ and $Y$ are said to be \textit{comonotonic} if $(X(\omega)-X(\omega'))(Y(\omega)-Y(\omega'))\geq0$ for all $\omega, \omega'\in\Omega$. We denote by  $X\laweq Y$ if the random variables $X$ and $Y$   have the same distribution.
For $x,y \in \R$, $x \vee y=\max\{x,y\}$ and $x \wedge y=\min\{x,y\}$.

In a reinsurance contract, the underlying risk $X$ faced by the insurer will be shared between the insurer and the reinsurer. 
Given an indemnity function, which is denoted by $I$, the risk $I(X) $ will be ceded to the reinsurer, while $R_I(X) = X - I(X) $ will be retained by the insurer. 
To avoid moral hazard, we define the set of all feasible indemnity functions as 
    \begin{equation}\label{I-set}
        \mathcal{I} = \left \{ 0\leq I(x)\leq x: I(x) \text{ is non-decreasing  and }0 \leq I(x) - I(y) \leq x - y \text{ for } 0 \leq y \leq x \right \}.
    \end{equation}
In exchange for risk transfer, the reinsurer can charge a reinsurance premium.
In a classical setup, the reinsurance premium is given by a premium principle, say $\pi$, which is a risk measure.
Consequently, the insurer and reinsurer's total losses become $R_I(X) + \pi (I(X)) $ and $ I(X) - \pi (I(X) )$, respectively.
In most reinsurance design problems discussed in the literature, 
the insurer adopts a risk measure $\rho $ to quantify her total risk exposure $  \rho \left ( R_I(X) + \pi (I(X)) \right )  $  for a given indemnity function $I \in \cI$, and 
then considers the optimization problem 
\begin{align}\label{Proj2_prob1}
           \min_{ I \in \mathcal{I} } \rho \left ( X-I (X) + \pi (I(X)) \right ). 
\end{align}
The problem \eqref{Proj2_prob1} and its variations have been intensively discussed in the insurance design literature. As this problem is framed from the insurer's perspective, the choice of the insurer's risk measure $\rho$ has attracted significant attention.
However, the choice of the reinsurance premium principle $\pi$, which represents the reinsurer's risk preference on quantifying her own part of loss, is a crucial component in the formulation of problem \eqref{Proj2_prob1} as well. 
Properties of the reinsurance premium principle may significantly affect the optimal solution to this problem. Relevant studies can be found in \cite{Young1999}, \cite{Kaluszke2001}, \cite{BT2009} and references therein.  

It is important to note that the reinsurer can leverage the reinsurance premium to influence the insurer's optimal strategy, potentially encouraging more favorable behavior. This influence becomes more pronounced when the reinsurer has greater flexibility to control the premium amount. In other words, transitioning from a constant premium to a variable premium scheme introduces additional randomness, further amplifying the reinsurer's ability to affect the insurer's decisions.

\subsection{Reward-and-penalty variable premium scheme}

Inspired by the rating plan presented in \cite{Meyers2004} and \cite{CCT2016}, we propose a ``reward-and-penalty" variable premium scheme for the reinsurer. This scheme defines the premium as a random variable at the outset of the policy period, with its realized value dependent on the insurer's performance. For a given ceded loss $Y \in L^1 $, the reinsurer first establishes a benchmark for the premium scheme using the expected-value premium principle. The reinsurer selects a positive risk loading $ \theta_0\geq0 $ and determines the \emph{``banchmark premium"} $ \pi_0 (Y) = (1+\theta_0) \E [Y ]$. 
In the second step, the reinsurer implements a performance-based premium plan based on the realized loss amount, $Y=y $.  If $y \leq \E [Y ]$, this outcome represents a ``reward scenario", and the reinsurer provides a discount on the premium. On the contrary,  if $y > \E [Y ] $, it triggers a ``penalty scenario", where the insurer needs to pay an extra penalization beyond the benchmark premium. The power of this ``reward-and-penalty" premium scheme is controlled by a scheme parameter $\delta$, determined by the reinsurer. Mathematically, the actual premium paid is expressed as $  \pi_0 (Y) + \delta (y - \E [Y ])$ with $\delta \in [0,1]$. This scheme parameter $\delta $ also signifies the increasing rate of the premium. 
Finally, the reinsurer determines the lower and upper bounds for the premium. To maintain consistent with the benchmark, the reinsurer applies the expected-value premium principle again. Two positive risk loadings,  $\theta_1\in[ 0, \theta_0]$ and $\theta_2\in[\theta_0, \infty]$, are selected to calculate the lower and upper bounds of the premium, given by $ \pi_1 (Y) = (1+\theta_1 ) \E [Y]$ and $ \pi_2 (Y) = (1+\theta_2 ) \E [Y]$, respectively. 
If the actual premium, $\pi_0 (Y) + \delta (y - \E [Y ]) $, falls below the lower bound $\pi_1(Y) $, the reinsurer charges the minimum payable premium $\pi_1 (Y) $. Conversely, if the actual premium exceeds the upper bound $\pi_2 (Y) $, the premium is capped at the maximum payable premium $\pi_2(Y)$.
\begin{definition}[Performance-based variable premium scheme]
Given a non-negative random loss $Y$ with finite mean, the ``reward-and-penalty" variable premium scheme on $Y$ is structured as follows
\begin{align}\label{premium}
    \pi_Y (y) & \triangleq \min \left\{ \max \left\{ \pi_0 (Y) + \delta(y - \mathbb E[Y]),  \pi_1 (Y) \right \}, \pi_2 (Y) \right\} \notag \\
    & = \min \left\{ \max \left\{ (1+\theta_0)\mathbb E[Y] + \delta(y - \mathbb E[Y]),  (1+\theta_1)\mathbb E[Y] \right \}, (1+\theta_2)\mathbb E[Y] \right\}, 
\end{align}
where $\delta \in [0,1]$ is the \emph{scheme parameter}, and $ \pi_i (Y) = (1+\theta_i) \E [Y] $ with risk loadings $\theta_i \geq 0 $, $i=0,1,2$.
\end{definition}
For a given 
$Y$, values of $\pi_i (Y) $, $i=0,1,2$, are based on the expected-value premium principle, and serve as key thresholds in the variable premium scheme.
Therefore, $\pi_Y$ in \eqref{premium} can be viewed as a tailor-made premium scheme function for the loss $Y$, with $\pi_Y(Y) $ being a random variable influenced by the distribution of $Y$, the scheme parameter $\delta$, and risk loading parameters $\theta_i$, $i=0,1,2$. 
The proposed variable premium scheme in \eqref{premium} can reduce to the standard expected-value premium principle under specific parameter settings. 
To demonstrate this, we can take either $\delta = 0 $ or $\theta_0=\theta_1=\theta_2$, which results in $\pi_Y(y) = (1+\theta_0)\mathbb E[Y]$ for all $y \geq 0 $.
To avoid redundant explanations, we assume $\delta> 0 $ and $\theta_1\leq \theta_0 <\theta_2$ in the following sections.

\begin{remark}
\cite{CCT2016} is a pioneering work that studies retrospective rating premium. They assume that the minimum premium is zero and formulate the premium scheme  as 
\begin{equation}\label{retro_CCT}
    \pi_{\text{RE}}(Y) = \min\left \{ B+L\cdot Y, (1+ \theta_{\max}\mathbb E[Y])\right \},
\end{equation}
where $\theta_{\max}\geq0$, and $B=B(Y) $, called as ``basic premium", is a free variable depending on $Y$ which is determined by solving $\mathbb E[\pi_{\text{RE}}(Y)] = (1+\theta_{\text{RE}})\mathbb E[Y]$ for some  $\theta_{\text{RE}}\geq0$. In our work, the variable premium defined in \eqref{premium} can also be written in the form of \eqref{retro_CCT} by taking $B = \pi_0(Y) - \delta\cdot \mathbb E[Y]$ and $\pi_1(Y) = 0$. 
However, our work assumes that the benchmark premium $\pi_0(Y) $ is directly given by the expected-value premium principle, and consequently, we have $\pi_Y(\mathbb E[Y]) = (1+\theta_0)\mathbb E[Y]$, $\theta_0\geq0$. In other words,  the value of $B$ in \cite{CCT2016} may vary for different choices of $Y$ with the same expectation. In contrast, within our framework, $\pi_0 (Y)$ remains fixed and the form of the premium does not change as long as the expectation of $Y$ is specified.
\end{remark}

Intuitively, when a risk measure $\pi$ is used as a reinsurance premium principle, it reflects the reinsurer's risk attitude toward the ceded loss. \cite{young2014premium} discusses the desirable properties of premium principles and lists popular ones used in the literature. 
In this work, we focus on the variable premium scheme, where $\pi_Y(Y) $ is a random variable. Although the properties of risk measures cannot be directly applied to $\pi_Y(Y)$, we can still discuss several desirable properties satisfied by this variable premium scheme. 

\begin{proposition}\label{prop-premium scheme}
Assume $\delta \in(0,1]$ and $ 0 \leq \theta_1\leq \theta_0 <\theta_2$ are given, and $Y \in L^1$ is a non-negative random loss. The reward-and-penalty variable premium scheme defined in \eqref{premium} satisfies the following properties. 
\begin{enumerate}[(i)]
\item 
The distribution of $\pi_Y(Y)$ depends only on the distribution of $Y$.
\item 
$\pi_Y(Y) \geq \E [Y ]$.
\item 
If $(1+\theta_2)\E[Y] \leq \esssup (Y)$, then $\pi_Y(Y) \leq \esssup (Y) $.
\item 
$\pi_{c\cdot Y} (c\cdot Y ) = c\cdot \pi_Y(Y) $ for all $c \geq 0 $. 
\item 
Let $Y_1$ and $Y_2$ be two non-negative random losses.  If $Y_1 $ is smaller than $Y_2$ in the sense of first-order stochastic dominance (FSD), denoted by $Y_1 \preceq_{FSD} Y_2 $,\footnote{Given two random losses $Y_1 $ and $Y_2$, the loss $Y_1 $ is said to be smaller than $Y_2$ in the sense of first-order stochastic dominance, denoted by $Y_1 \preceq_{FSD} Y_2 $, if $F_{Y_1}^{-1} (p) \leq F_{Y_2}^{-1} (p) $ for all $p \in [0,1]$. } then $ \pi_{Y_1} (Y_1 ) \preceq_{FSD} \pi_{Y_2}(Y_2) $.
\item 
Functions $\pi_Y(y) $ and $y- \pi_Y(y) $ are continuous and non-decreasing in $y \geq 0$, and moreover, random variables $Y$, $\pi_Y(Y)$ and $Y- \pi_Y(Y)$ are comonotonic. 
\end{enumerate}
\end{proposition}
\begin{proof}
\begin{enumerate}[(i)]
\item Since $\pi_i(Y) =(1+\theta_i) \E[Y]$, $i=0,1,2,$ are law-invariant, it is easy to see from \eqref{premium} that the function $\pi_Y(y) $ is determined by the distribution of $Y$. Hence, the distribution of $\pi_Y(Y)$ depends only on the distribution of $Y$.
\item 
Since $\theta_1 \geq 0 $, the equation \eqref{premium} implies 
\begin{align}\label{prop1-eq1}
     \pi_{Y} (y)& = \begin{cases} (1+\theta_1) \E [Y], & \text{ if }Y \leq \left ( \frac{\theta_1 - \theta_0+ \delta }{\delta } \E [Y]\right )_+, \\ 
     (1+\theta_0) \E [Y] + \delta (y - \E [Y] ) , & \text{ if }\left ( \frac{\theta_1 - \theta_0+ \delta }{\delta }\E [Y]\right )_+ < Y < \frac{\theta_2 -\theta_0 + \delta }{\delta }\E[Y], \\
     (1+\theta_2) \E[Y], & \text{ if }y \geq \frac{\theta_2 - \theta_0+ \delta }{\delta }\E[Y] . \end{cases}
\end{align}
It then follows that $\pi_{Y} (Y) \geq  (1+\theta_1) \E [Y] \geq \E [Y]$.
\item 
If $(1+\theta_2)\mathbb E[Y] \leq \esssup (Y)$, then  \eqref{prop1-eq1} implies $\pi_{Y} (Y)\leq (1+\theta_2) \E [Y] \leq \esssup (Y)  $.
\item 
Take a constant $c \geq 0 $. It is easy to check that
\begin{align*}
       \pi_{c Y} (cY )
       & =  \min \left\{ \max \left\{ (1+\theta_0)\mathbb E[cY] + \delta(cY- \mathbb E[cY]),  (1+\theta_1)\mathbb E[cY] \right \}, (1+\theta_2)\mathbb E[cY] \right\}\\
       & = \min \left\{ c \cdot \max \left\{ (1+\theta_0)\mathbb E[Y] + \delta(Y- \mathbb E[Y]),   (1+\theta_1)\mathbb E[Y] \right \}, c(1+\theta_2)\mathbb E[Y] \right\}\\
       & = c \min \left\{ \max \left\{ (1+\theta_0)\mathbb E[Y] + \delta(Y- \mathbb E[Y]),   (1+\theta_1)\mathbb E[Y] \right \}, (1+\theta_2)\mathbb E[Y] \right\} \\
       & = c \cdot \pi_Y(Y) .
\end{align*}
\item
Note that $Y_1 \preceq_{FSD} Y_2 $ implies $ \Pr (Y_1 \leq z ) \geq \Pr (Y_2 \geq z ) $ for all $z \geq 0 $, and thus $\mathbb E[Y_1]\leq \mathbb E[Y_2]$. 
The premium $\pi_{Y_1} (Y_1)$ and $\pi_{Y_2} (Y_2)$ are non-negative random variables structured as in \eqref{prop1-eq1}.
\begin{enumerate}
\item
For $ 0\leq z < (1+\theta_1) \mathbb E[Y_2]  $, if any, we have $\Pr ( \pi_{Y_2} (Y_2) \leq z ) = 0 \leq \Pr ( \pi_{Y_1} (Y_1) \leq z )$.
\item 
For $  (1+\theta_1) \mathbb E[Y_2]  \leq z < (1+\theta_2 ) \mathbb E[Y_1] $, if any, we have
\begin{align*}
       \Pr ( \pi_{Y_2} (Y_2) \leq z ) & = \Pr \left ( (1+\theta_0) \mathbb E[Y_2] + \delta (Y_2 - \mathbb E[Y_2] ) \leq z  \right )\\
       & =  \Pr \left ( Y_2 \leq \frac{1}{\delta} ( z - (1+\theta_0- \delta )\mathbb E[Y_2] ) \right ) \\
       & \leq  \Pr \left ( Y_2 \leq \frac{1}{\delta} ( z - (1+\theta_0- \delta )\mathbb E[Y_1] ) \right ) \\
       & \leq  \Pr \left ( Y_1 \leq \frac{1}{\delta} ( z - (1+\theta_0- \delta )\mathbb E[Y_1] ) \right ) 
       = \Pr ( \pi_{Y_1} (Y_1) \leq z ) 
\end{align*}
\item 
For $ z\geq (1+\theta_2 ) \mathbb E[Y_1] $, we have $ \Pr ( \pi_{Y_2} (Y_2) \leq z ) \leq 1 = \Pr ( \pi_{Y_1} (Y_1) \leq z ) $. 
\end{enumerate}
Combining three cases, we conclude that $ \Pr ( \pi_{Y_2} (Y_2) \leq z ) \leq  \Pr ( \pi_{Y_1} (Y_1) \leq z ) $ for all $ z \geq 0 $. It follows that $ \pi_{Y_1} (Y_1 ) \preceq_{FSD} \pi_{Y_2}(Y_2)$. 
\item 
From equation \eqref{prop1-eq1}, since $\delta \in (0,1]$, it follows that both $\pi_Y(y) $ and $y-\pi_Y(y) $ are continuous and non-decreasing functions of $y$.
Thus, $Y$, $\pi_Y(Y)$ and $Y- \pi_Y(Y)$ are comonotonic. 
\qedhere
\end{enumerate}
\end{proof}

We can compare the properties outlined in Proposition \ref{prop-premium scheme} with those properties commonly associated with risk measures for premium principles. For a comprehensive discussion on the economic interpretation of these properties, see \cite{young2014premium}.

\begin{table}[ht]
\centering
\label{tab: var}

\begin{tabular}{c|lll}
\hline
Proposition \ref{prop-premium scheme} &                   
 \multicolumn{2}{c}{Properties for a premium principle $\pi$} \\ \hline 
Property (1) & ``Law-invariance": & $\pi(Y_1)=\pi(Y_2) $ if $Y_1\laweq Y_2 $\\ \hline
Property (2) & ``Risk loading": & $\pi(Y)\geq \E[Y] $ for all $Y$\\ \hline
Property (3) & ``No rip-off": & $\pi(Y)\leq \esssup Y $ for all $Y$\\ \hline
Property (4) & ``Homogeneity": & $\pi(cY) = c \pi(Y) $ for all $Y$ and constant $c > 0 $ \\ \hline
Property (5) & ``Preserving FSD": & $\pi(Y_1) \leq \pi(Y_2) $ if $Y_1 \preceq_{FSD} Y_2 $ \\ \hline
\end{tabular}
\end{table}

The property (6) in Proposition \ref{prop-premium scheme} further implies that functions $\pi_Y(y)$ and $y - \pi_Y(y) $ exhibit 1-Lipschitz continuity.
This aligns with the condition for a feasible indemnity function as defined in \eqref{I-set}. 
Suppose that the distribution of the underlying loss $X$ faced by the insurer is known in the optimal reinsurance problem, then the distribution of the reinsurer's ceded loss $I(X) $ is determined once the indemnity function $I \in \cI $ is specified. 
The 1-Lipschitz continuity condition on $I $ ensures that $X$, $I(X)$ and $X- I(X) $ are monotonic random variables. This comonotonic property, also referred to as the ``no-sabotage" condition, helps to mitigate moral hazard issues in the reinsurance policy.
When the reinsurer applies the reward-and-penalty variable premium scheme in \eqref{premium}, the premium variable then becomes 
\begin{equation*}
    \Pi_{I}(X) \triangleq \pi_{I(X)}(I(X))= \min \left\{ \max \left\{ \pi_0 (I(X))+ \delta(I(X) - \mathbb E[I(X)]), \pi_1 (I(X)) \right \},  \pi_2 (I(X)) \right\},
\end{equation*}
where $\pi_i (I(X)) =  (1+\theta_i)\mathbb E[I(X)]$, $i=0,1,2$.
The insurer and reinsurer's total losses become $ X- I(X) + \Pi_{I} (X) $ and $ I(X) - \Pi_{I } (X) $, respectively. 
Since $I \in \cI$ ensures that $X$, $I(X)$, and $X - I(X)$ are comonotonic, Property (6) in Proposition \ref{prop-premium scheme} further implies that $X$, $X- I(X) + \Pi_{I} (X) $ and $ I(X) - \Pi_{I} (X) $ are comonotonic random variables. 
As a result, the ``no-sabotage" condition continues to hold. 

In the remainder of the paper, we use the variable premium scheme to formulate optimization problems, as specified in the following assumption.

\begin{assumption}\label{assumption1}
Assume $\delta \in(0,1]$,  $ (\theta_0 - \delta) \vee 0\leq \theta_1\leq \theta_0 <\theta_2 <\infty $, and the non-negative underlying loss $X \in L^1 $ is given. 
For any indemnity function $I \in \cI$,  the reinsurer employs the following ``reward-and-penalty" variable premium scheme for the ceded loss $I(X) $:
\begin{align}\label{premium_I}
    \Pi_{I}(x) &= \min \left\{ \max \left\{ \pi_0 (I(X))+ \delta(I(x) - \mathbb E[I(X)]), \pi_1 (I(X)) \right \},  \pi_2 (I(X)) \right\} \notag \\
    &=
\begin{cases}
            (1+\theta_1)\mathbb E[I(X)], & \text{ if }  0 \leq I(x) \leq d_I,\\
            (1+\theta_0)\mathbb E[I(X)] + \delta ( I(x) - \mathbb E[I(X)] ) , & \text{ if } d_I \leq I(x) \leq u_I,\\
            (1+\theta_2)\mathbb E[I(X)], & \text{ if } I(x) \geq u_I,
        \end{cases}
\end{align}
where $d_I = \frac{\theta_1 - \theta_0 + \delta}{\delta}\mathbb E[I(X)]$, $u_I = \frac{\theta_2 - \theta_0 + \delta}{\delta}\mathbb E[I(X)]$, and $\pi_i (I(X)) = (1+\theta_i) \E [I(X) ]$ for $i=0,1,2$. 
\end{assumption}

The condition $ (\theta_0 - \delta) \vee 0\leq \theta_1$ in Assumption \ref{assumption1} ensures that $d_I \geq 0$ to prevent trivial cases where the lower threshold for the premium calculation would be negative or meaningless.

\section{Insurer's optimal reinsurance strategy}\label{sec:insurer's opt prob}

In this section, we solve the insurer's optimization problem:  
\begin{align}\label{insurer's prob with variable prem}
       \min_{I \in \cI } \rho \left ( T_I(X)  \right ) , ~~~~~\text{ where }  T_I(X) \triangleq R_I(X) + \Pi_{I}(X) \text{ for any }  I \in \cI.
\end{align}
Under Assumption \ref{assumption1}, the premium $ \Pi_{I}(X)$ is again a random variable. Since the amount of the premium paid depends on the realized indemnity payment amount, which cannot be observed at the beginning of the policy period, the insurer needs to consider the influence of a given indemnity function $I$ and the distribution of $X$ on the variable premium. 
In addition to the variable premium scheme, the choice of the risk measure $\rho$ significantly influences the solution to the problem \eqref{insurer's prob with variable prem}. 
To begin, we impose mild conditions on $\rho $ and characterize the optimal reinsurance policy for the insurer. We then focus on a distortion risk measure $\rho$, which enables us to further simplify the structure of solutions to \eqref{insurer's prob with variable prem}.

\subsection{Optimal reinsurance design with a general risk measure}\label{sec:opt sol with general RM}

This section presents the main result of the paper, where the infinite-dimensional problem \eqref{insurer's prob with variable prem} is reduced to a finite-dimension optimization problem, provided that the risk measure $\rho$ meets the following two conditions:
\begin{enumerate}[(i)]
\item Law-invariance: $\rho(X) = \rho (Y) $ if $X \laweq Y $;
\item Preserving the convex order\footnote{Given two random losses $Y_1$ and $Y_2$, the loss  $Y_1$ is said to be smaller than $Y_2$ in the convex order, denoted by $Y_1 \preceq_{cx} Y_2 $, if $\E [h(Y_1 ) ] \leq \E [h(Y_2) ]$ for all convex functions $h$ such that the expectations exist. }: $\rho(X) \leq \rho(Y) $ if $X \preceq_{cx} Y $.
\end{enumerate}

To proceed, define 
\begin{align}\label{cI-tilde}
\tilde {\mathcal I} = \mathcal I_1\cup\mathcal I_2, 
\end{align}
where
\begin{align*}
\mathcal I_1  &=\left \{I(x) = (x-d_1)_+ - (x-d_1-d_I)_+ + (x-d_2)_+  \left \vert \begin{aligned} & d_I =  \frac{\theta_1 - \theta_0 + \delta}{\delta}\mathbb E[I(X)] \\& \text{ and }0\leq d_1\leq d_1 + d_I \leq d_2 \leq\infty \end{aligned} \right . \right \} ,
\end{align*}
and 
\begin{align*}
\mathcal I_2 &= \left \{ I(x) = (x-d_1)_+ - (x-d_1-u_I)_+ + (x-d_2)_+  \left \vert \begin{aligned} & u_I =  \frac{\theta_2 - \theta_0 + \delta}{\delta}\mathbb E[I(X)] \\& \text{ and } 0\leq d_1\leq d_1 + u_I \leq d_2 \leq\infty \end{aligned} \right . \right \}.
\end{align*}

\begin{figure}[htbp]
    \centering
    \begin{subfigure}[b]{0.49\textwidth}
        \centering
        \includegraphics[width=\textwidth]{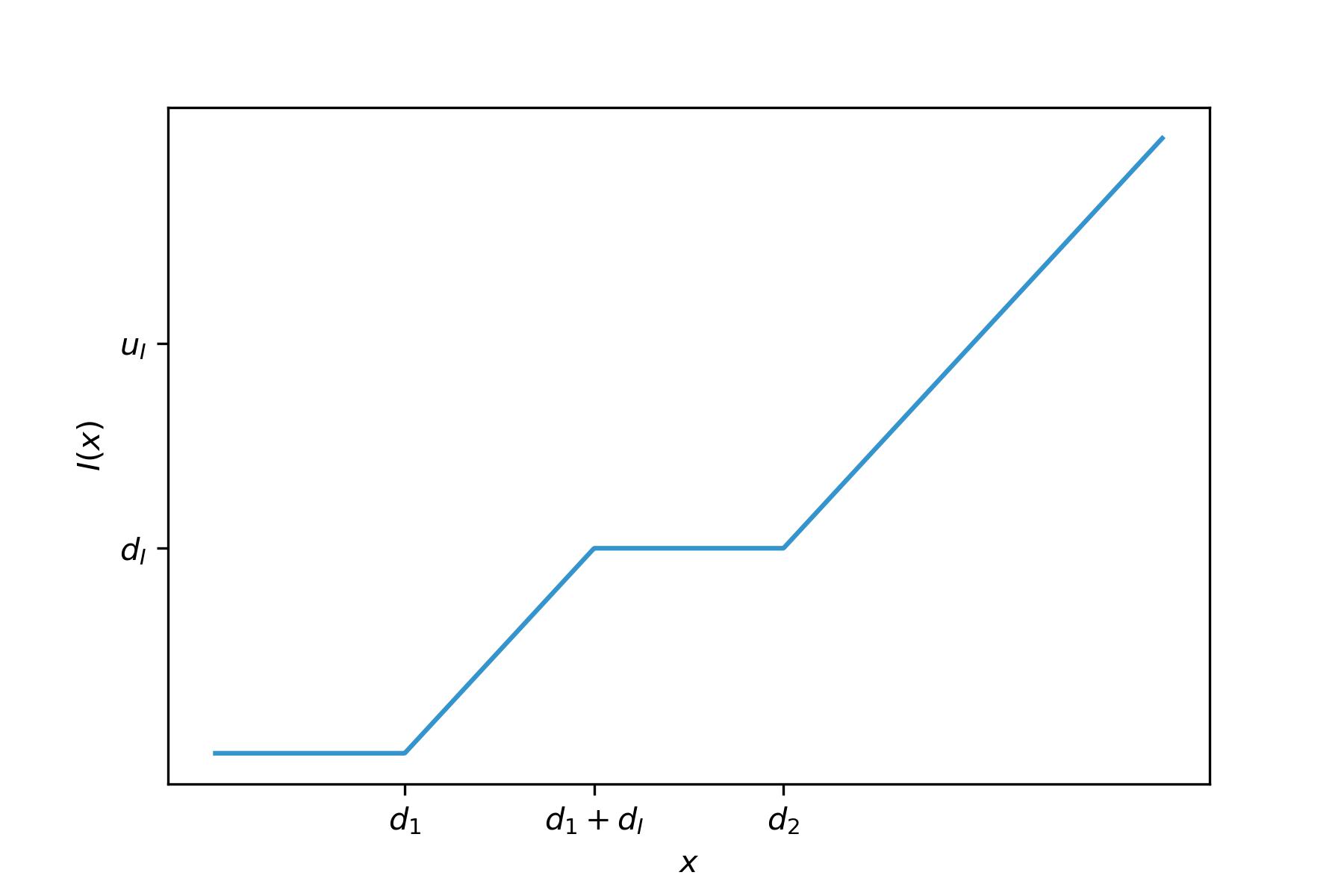}
        \caption{}
        \label{fig:I_class1}
    \end{subfigure}
    \hfill
    \begin{subfigure}[b]{0.49\textwidth}
        \centering
        \includegraphics[width=\textwidth]{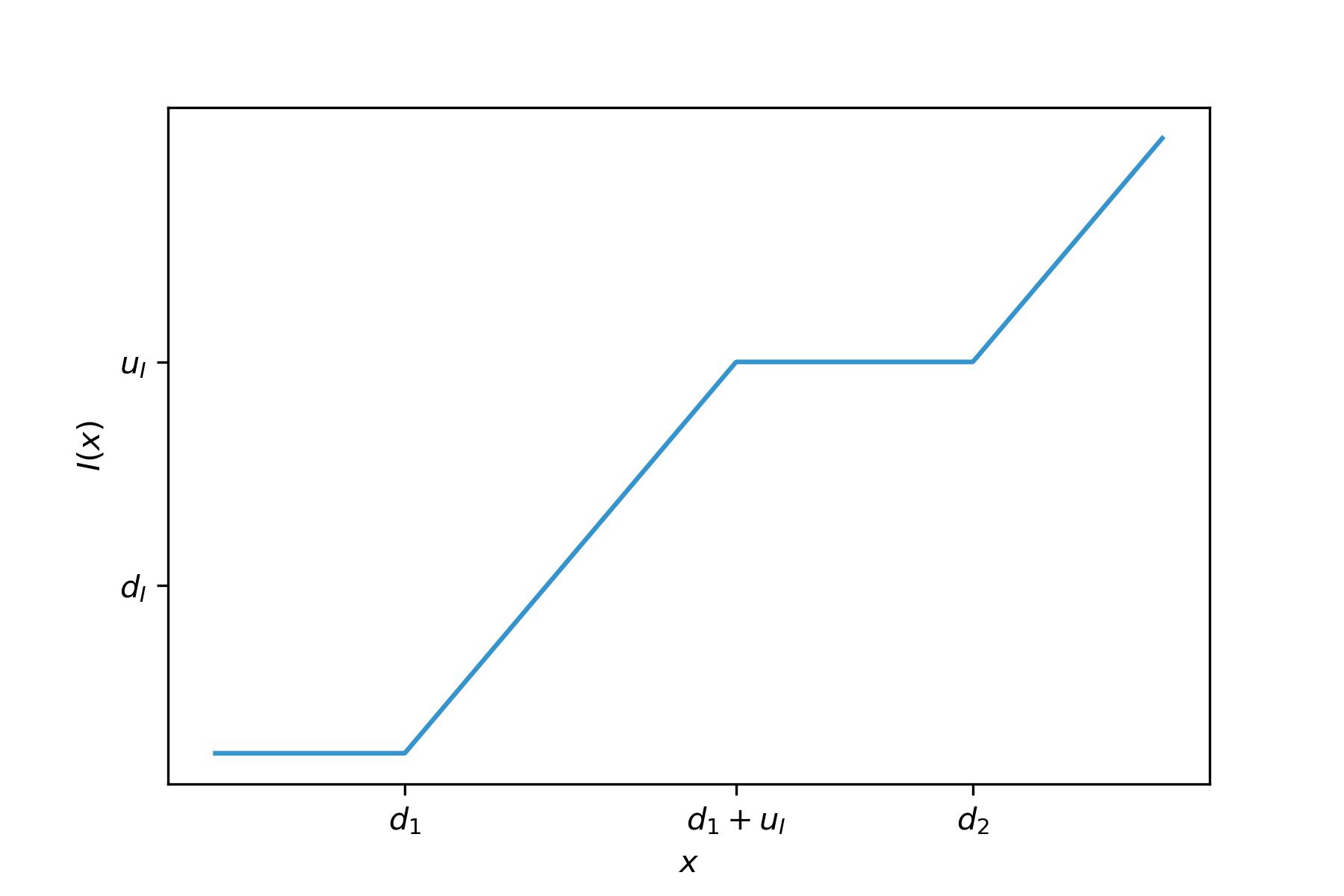}
        \caption{}
        \label{fig:I_class2}
    \end{subfigure}
    \caption{Indemnity functions $I(x)$, where (a) $I(x)\in\mathcal I_1$, or (b) $I(x)\in\mathcal I_2$.}
    \label{Fig:I_class}
\end{figure}

Note that stop-loss functions belong to $\mathcal I_1$ and $\mathcal I_2$. To see this, we first fix a real value $a$ such that $0\leq a\leq \mathbb E[X]$. There exists $\tilde d\geq 0$ such that 
$\mathbb E[(X-\tilde d)_+] = \int_{\tilde d}^\infty S(x)\d x = a.$ 
Then we can take $d_1 = \tilde{d} $, $d_2 = d_1 + \frac{\theta_1 - \theta_0+ \delta }{\delta }a= d_1 + d_I $, and it is easy to see 
$$I(x) = (x - d_1 )_+ - (x- d_1 -d_I)_+ + (x - d_2 ) = (x - d_1 )_+ = (x - \tilde{d} )_+ \in \mathcal{I}_1 .$$
Therefore, $\mathcal{I}_1 \neq \emptyset $. Similarly, we can argue that $(x - \tilde{d} )_+ \in \mathcal{I}_2 $ and $ \mathcal{I}_2 \neq \emptyset$. Consequently, the set $\tilde{\mathcal{I}} $ is well-defined.

The shapes of the indemnity functions in $\mathcal I_1$ and $\mathcal I_2$ are illustrated in Figure \ref{Fig:I_class}. The elements in $\tilde {\mathcal I}$ represents two-layer insurance policies. We will demonstrate that the optimal indemnity function can be obtained within the set $\tilde {\mathcal I}$. 
To proceed, we first take an intermediate step by verifying that the optimal solution to problem \eqref{insurer's prob with variable prem} belongs to the following three-layer policy set.
     \begin{align}\label{S3}
    \mathcal S_3 = \Big \{ & \, f(x) = (x-a)_+ - (x-b)_+ +(x-c)_+ - (x-d)_+ + (x-e)_+, \, x \geq 0, \notag \\
     & \, \text{ where } 0\leq a\leq b = a+d_I\leq c\leq d = c+ u_I-d_I\leq e \Big\}.
    \end{align}

\begin{lemma}\label{lem:I_opt_3layer}
   Let Assumption \ref{assumption1} hold. Assume that $\rho$ is a law-invariant risk measure that preserves the convex order. For any $I\in \cI$,  there exists $f_I \in \mathcal{S}_3$ such that $\rho \left ( T_{f_I}(X)  \right ) \leq \rho \left ( T_I(X)  \right )$. 
   Furthermore, solving problem \eqref{insurer's prob with variable prem} reduces to solving the following problem
    \begin{align}\label{I_opt_3layer}
        \min_{I\in\mathcal{S}_3}\rho \left ( T_{I}(X)  \right ) .
    \end{align}   
\end{lemma}

The proof of Lemma \ref{lem:I_opt_3layer} is provided in \ref{app:proof in sec3.1}. This lemma enables us to focus on the optimal policy within the three-layer policies defined in $\mathcal{S}_3$. Utilizing this result, we can further refine the search for optimal policies. 
The proof of the following theorem is also presented in \ref{app:proof in sec3.1}. 

\begin{theorem}\label{Proj2_thm1}
    Let Assumption \ref{assumption1} hold. Assume that $\rho$ is a law-invariant risk measure that preserves the convex order. For any $I\in \cI$,  there exists $f_I \in \tilde{\cI}$ such that $\rho \left ( T_{f_I}(X) \right ) \leq \rho \left ( T_I(X)  \right )$. Furthermore, to solve problem \eqref{insurer's prob with variable prem}, it is sufficient to solve the following problem:
    \begin{align}\label{I_opt_2layer}
        \min_{I\in\tilde{\mathcal{I}}}\rho \left ( T_{I}(X) \right ) .
    \end{align}
\end{theorem}

By Theorem \ref{Proj2_thm1}, the problem \eqref{insurer's prob with variable prem} is equivalent to problem \eqref{I_opt_2layer}, and the optimal indemnify function belongs to  $\tilde{\mathcal I}$, which is indexed by a finite number of parameters. 

It is worth highlighting that Theorem \ref{Proj2_thm1} requires nothing but law-invariance and preservation of convex order on the risk measure $\rho$. These assumptions encompass a wide range of risk measures. However, without specifying the exact form of $\rho$, identifying the optimal solution within $\tilde{\mathcal{I}}$ becomes mathematically challenging.
To further investigate the insurer's optimal solution, we impose additional conditions on the risk measure $\rho$ in the following section.

\subsection{Optimal reinsurance design with a distortion risk measure}\label{proj2-sec2}

In this section, we consider the family of distortion risk measures. A risk measure $\rho $ is referred to as a \emph{distortion risk measure} if it can be expressed in the following way:
\begin{align}\label{distortion}
\rho^g(X) = \int_0^\infty g (S_X(x)) \d x - \int_{-\infty }^0 \left [ 1 - g (S_X(x)) \right ] \d x ,
\end{align}
where $ g : [0,1] \rightarrow [0,1]$ is a distortion function, characterized as a non-decreasing function with $g(0) = 0 $ and $g(1) = 1 $.
In this work, we assume that the value of $\rho^g(X) $ in \eqref{distortion} is finite whenever it is used.

Commonly used  distortion risk measures include the
(left) \emph{Value-at-Risk} (VaR) and the \emph{Tail Value-at-Risk} (TVaR).
Given a confidence level $\alpha\in (0,1)$, $\VaR_\alpha$ is defined  as
$\VaR_\alpha (X) =F_X^{-1} ( 1-\alpha) $, $X\in L^0 $,
with the distortion function $ g(p) = \id_{( \alpha, 1 ] }(p)$;
and $\TVaR_\alpha$ is  defined as
\begin{align}\label{def_distortion_tvar}
           \TVaR_\alpha (X) = \frac{1}{\alpha}\int_0^\alpha \VaR_u ( X) \d u = \int_0^\infty g(S_X(x))\d x ,~~~X\in L^1 ,
\end{align}
with the distortion function 
\begin{equation} \label{proj2-TVaR-distortion}
    g(p) = \left\{
    \begin{aligned}
        & \frac{p}{\alpha}, &p\leq \alpha,\\
        & 1, & p > \alpha.
    \end{aligned}
    \right.
\end{equation}
It is well-known that $\TVaR_\alpha$ is a coherent risk measure with a concave distortion function $g(p)$ in \eqref{proj2-TVaR-distortion}, whereas $\VaR_\alpha$ is not coherent with a non-concave distortion function $\id_{(\alpha, 1]}$.
According to Theorem 3 of \cite{WWW2020}, a distortion risk measure $\rho^g$ is convex order consistent if and only if it is convex,\footnote{A risk measure $\rho$ satisfies the convexity property if $\rho(\lambda X + (1-\lambda)Y ) \leq \lambda \rho(X) + (1-\lambda) \rho(Y) $ for any random variables $X$ and $Y$ and constant $\lambda \in [0,1]$.} which is  further equivalent to the concavity of  its distortion function $g$.
Additionally, by its definition in \eqref{distortion}, $\rho^g$ is law-invariant and homogeneous. Therefore, taking a distortion risk measure $\rho^g $ that is convex order consistent is equivalent to assuming that $\rho^g $ is coherent.

The family of distortion risk measures is widely applied in the study of insurance and reinsurance optimization problems. It encompasses popular risk measures, such as VaR and TVaR, which are used to quantify the insurer/reinsurer's risk exposure, as well as premium principles like the expected-value premium principle and Wang's premium principle. 
Moreover, its expression in \eqref{distortion} provides a useful mathematical framework for tracking optimal solutions.

\begin{theorem}\label{proj2_thm2}
    Let $\rho = \rho^g$ be a coherent distortion risk measure and Assumption \ref{assumption1} hold. Then the optimal indemnity function $I^*$ of problem $\min_{I\in\tilde{\mathcal I}}\rho \left ( T_I(X) \right )$ is contained in the set $\mathcal I_1$. That is,
    \begin{equation}\label{Proj2_dedimension2}
        \min_{I\in\tilde{\mathcal I}}\rho \left ( T_I(X) \right ) = \min_{I\in\mathcal I_1}\rho \left ( T_I(X) \right ).
    \end{equation}
\end{theorem}
\begin{proof}
    Let $I(x) = (x-d_1)_+ - (x - d_1 - u_I)_+ +(x-d_2)_+$ with $0\leq d_1\leq d_1 + u_I\leq d_2$, which is an element in the policy set $\mathcal I_2$. Given $\mathbb E[I(X)] = a$, then $d_I = \frac{\theta_1-\theta_0+\delta}{\delta}a$ and $u_I = \frac{\theta_2-\theta_0+\delta}{\delta}a$. 
    By \eqref{premium_I}, the survival function of $\Pi_{I}(X)$ is given by
    \begin{equation}\label{S_pi_2}
        S_{\Pi_{I}(X)} (x)  = 
        \begin{cases}
            1, &x<(1+\theta_1)a,\\
            S_X\left (\frac{x}{\delta}-\frac{a(1+\theta_0-\delta)}{\delta} + d_1 \right ), & (1+\theta_1)a \leq x < (1+\theta_2)a,\\
            0, &x\geq(1+\theta_2)a,
        \end{cases}
    \end{equation}
    from which we can obtain that
        $$\rho^g(\Pi_{I}(X)) = \int_0^\infty g(S_{\Pi_{I}(X)}(x)) \d x = (1+\theta_1)a + \delta \int_{d_1+d_I}^{d_1 + u_I}g(S_X(x))\d x.$$
    In addition,
        $\rho^g(R_{I}(X)) = \int_0^{d_1}g(S_X(x))\d x + \int_{d_1+u_I}^{d_2}g(S_X(x))\d x.$
    Note that $ R_I(x) $ and $\Pi_{I}(x)$ are both increasing functions of $x$, implying that the retained loss $R_I(X) $ and the variable premium scheme $\Pi_{I}(X)$ are comonotonic random variables. By the comonotonic additivity property of $ \rho^g $, we have
    \begin{equation*}
    \begin{aligned}
        \rho^g(T_{I}(X)) 
        = &\rho^g(R_{I}(X)) + \rho^g(\Pi_{I}(X)) \\    
        = & \int_0^{d_1}g(S_X(x))\d x + \int_{d_1+u_I}^{d_2}g(S_X(x))\d x
           +\delta \int_{d_1+d_I}^{d_1+u_I}g(S_X(x))\d x + (1+\theta_1)a.
    \end{aligned}
    \end{equation*}
    It then follows that
    \begin{align*}
        \begin{cases} \frac{\partial\rho^g(T_I(X))}{\partial d_1}  = g(S_X(d_1)) - \delta g(S_X(d_1 + d_I)) - (1-\delta)g(S_X(d_1+u_I)),\\
         \frac{\partial\rho^g(T_I(X))}{\partial d_2} = g(S_X(d_2)).\end{cases}
    \end{align*}
    Since $\mathbb E[I(X)] = \int_{d_1}^{d_1+u_I}S_X(x)\d x + \int_{d_2}^{\infty}S_X(x) \d x$, by Implicit Function Theorem, we have 
        $$\frac{\d d_2}{\d d_1} = - \frac{\partial \mathbb E[I(X)]/\partial d_1}{\partial \mathbb E[I(X)]/\partial d_2} = \frac{S_X(d_1 + u_I) - S_X(d_1)}{S_X(d_2)}.$$
    Thus, we can get
    \begin{equation}\label{drho_dd1}
    \begin{aligned}
        \frac{\d}{\d d_1} \rho^g(T_{I}(X)) 
        = & \frac{\partial}{\partial d_1}\rho^g(T_{I}(X)) + \frac{\d d_2}{\d d_1}\left ( \frac{\partial}{\partial d_2}\rho^g(T_{I}(X)) \right )\\
        = & g(S_X(d_1)) - \delta g(S_X(d_1 + d_I)) - (1-\delta)g(S_X(d_1+u_I))\\
        & + g(S_X(d_2))\frac{S_X(d_1 + u_I) - S_X(d_1)}{S_X(d_2)}\\
        \leq & [S_X(d_1) - S_X(d_1 + u_I)]\left[ \frac{g(S_X(d_1) - g(S_X(d_1+u_I))}{S_X(d_1) - S_X(d_1+u_I)} - \frac{g(S_X(d_2))}{S_X(d_2)} \right]\\
        \leq & 0.
    \end{aligned}
    \end{equation}   
    Here, the first inequality follows from the fact that $g(S_X(d_1+d_I))\geq g(S_X(d_1+u_I))$, while the second inequality is the result of the concavity of $g$. Thus, $\rho^g(T_{I}(X))$ is a decreasing function of $d_1$. Thus, the minimum value of $\rho^g(T_{I}(X))$ is achieved when $d_1 = d_2-u_I$. This indicates that the stop-loss policy $I^*(x) = (x - d_1^*)_+$ is the optimal solution to the problem $\min_{I\in\mathcal I_2, \mathbb E[I(X)] = a}\rho \left ( T_I(X) \right )$, with the optimal parameter $d_1^*$ determined by the condition $\mathbb E[I^*(X) ]= a$. Noting that stop-loss functions are also elements of $\mathcal I_1$, it follows that, for a given $a$, $\min_{I\in\mathcal I_1, \mathbb E[I(X)] = a}\rho \left ( T_I(X) \right )$ is no larger than $\min_{I\in\mathcal I_2, \mathbb E[I(X)] = a}\rho \left ( T_I(X) \right )$. Therefore, we conclude that \eqref{Proj2_dedimension2} holds, which completes the proof.
\end{proof}

Recall from Theorem \ref{Proj2_thm1} that the range of the optimal solution to problem \eqref{Proj2_prob1} is reduced to the set $\tilde{\cI} $. By its definition in \eqref{cI-tilde}, there are two types of reinsurance structure, denoted as $\cI_1 $ and $\cI_2$. 
Theorem \ref{proj2_thm2} further refines the range of optimal solutions from $\tilde{\cI}$ to $\cI_1$ when a coherent distortion risk measure is employed.  
According to \eqref{cI-tilde}, any $ I \in \cI_1$ takes the form $I(x) = (x-d_1)_+ - (x-d_1-d_I)_+ + (x-d_2)_+$, where $d_I =  \frac{\theta_1 - \theta_0 + \delta}{\delta}\mathbb E[I(X)]$ and $0\leq d_1\leq d_1 + d_I \leq d_2 \leq\infty$.
Mathematically, the set $\cI_1 $ can be partitioned into layers based on the expected value $\E [I(X) ]$, i.e.,  $ \cI_1 = \bigcup_{0 \leq a \leq \E[X]} \{I \in \cI_1: \E [I(X)] = a \}$. 
Within each layer, the expected value $\E [I(X)] = a$ is fixed.
Consequently,  the parameter $d_I$ remains constant for all indemnity functions within this layer. Meanwhile, the values of $d_1 $ and $d_2$ are restricted by the expected value such that 
$ \E [I(X) ] = \left (\int_{d_1}^{d_1 +d_I} + \int_{d_2}^\infty \right ) S_X(x) \d x = a .$
Therefore, each layer $\{I \in \cI_1: \E [I(X)] = a \}$ only has one free parameter.
Consequently, the problem \eqref{Proj2_dedimension2} can be reformulated as a two-step minimization problem
\begin{align*}
\min_{0 \leq a \leq \E [X] } \left \{ \min_{I \in \cI_1, \E [I(X)] = a } \rho \left ( T_I(X) \right ) \right \} , 
\end{align*}
where each inner minimization problem involves a single variable, making it feasible to be solved numerically.

In the following proposition, we take $\rho = $TVaR$_\alpha$ in the above problem and derive the closed-form solution to the inner optimization problem. Specifically, we solve the following problem:
\begin{equation}\label{Proj2_prob_I1_a}
    \min_{I\in\mathcal I_1, \mathbb E[I(X)] = a}\TVaR_\alpha \left ( T_I(X) \right ) .
\end{equation}

\begin{proposition}\label{lemma_opt_parameters_1}
    Let $\rho = \TVaR_\alpha$ for some $0 <\alpha < 1 $, $ 0< a \leq\mathbb E[X] $, and Assumption \ref{assumption1} hold. Define $d_I =a( \theta_1 - \theta_0 + \delta)/ \delta $, $u_I =a( \theta_2 - \theta_0 + \delta)/ \delta $, and let $\tilde d$ be the solution to $\mathbb E[(X-d)_+] = a$.  If $\frac{S_X(d+u_I-d_I)}{S_X(d)}$ is an increasing function of $d$,      
    then the optimal indemnity function for problem \eqref{Proj2_prob_I1_a} is given by
    \begin{align}\label{I_opt_fixa_cvar}
    I^*(x) = \begin{cases}  (x-\bar{d}_1)_+ - (x - \bar{d}_1 - d_I)_+ +(x-\bar{d}_2)_+, & \text{ if }\bar{d}_1 < \tilde{d}, \\ 
    (x - \tilde{d})_+ , & \text{ if } \bar{d}_1 \geq \tilde{d} ,\end{cases}
    \end{align}
    where the pair $ (\bar{d}_1, \bar{d}_2) $ is the solution to the following equation system 
    \begin{align}\label{proj2_op_parameters}
       \left \{
       \begin{aligned}    
       & \frac{\alpha - S_X(d_1+d_I)}{S_X(d_1) - S_X(d_1+d_I)}-\delta\frac{S_X(d_2+u_I-d_I)}{S_X(d_2)}+\delta-1 = 0, \\
       &  \int_{d_1}^{d_1 + d_I}S_X(x)\d x + \int_{d_2}^\infty S_X(x)\d x = a. \end{aligned}\right .
    \end{align}
\end{proposition}

\begin{proof}
    Let $I(x) = (x-d_1)_+ - (x - d_1 - d_I)_+ +(x-d_2)_+$ with $0\leq d_1\leq d_1 + d_I\leq d_2$, which is an element in the policy set $\mathcal I_1$. The domain of $d_1$ is clearly $[0,\tilde d]$. Given that $\mathbb E[I(X)] = a$, we have $d_I = \frac{\theta_1-\theta_0+\delta}{\delta}a$ and $u_I = \frac{\theta_2-\theta_0+\delta}{\delta}a$.  
    
    Since $\mathbb E[I(X)] = \int_{d_1}^{d_1+d_I}S_X(x)dx + \int_{d_2}^{\infty}S_X(x)dx = a$, by Implicit Function Theorem, $d_2$ can be treated as an implicit function of $d_1$, and
    \begin{equation}\label{dd2_dd1}
        \frac{\d d_2}{\d d_1} = - \frac{\partial \mathbb E[I(X)]/\partial d_1}{\partial \mathbb E[I(X)]/\partial d_2} = \frac{S_X(d_1 + d_I) - S_X(d_1)}{S_X(d_2)}.
    \end{equation}
    By \eqref{premium_I}, the survival function of $\Pi_{I}(X)$ is formulated as
    \begin{equation}\label{S_pi_1}
        S_{\Pi_{I}(X)}(x) =
        \begin{cases}
            1, &x<(1+\theta_1)a,\\
            S_X(\frac{x}{\delta}-\frac{a(1+\theta_1)}{\delta} + d_2), & (1+\theta_1)a \leq x < (1+\theta_2)a,\\
            0, &x\geq(1+\theta_2)a.
        \end{cases}
    \end{equation}
    Denote $g$ the distortion function of $\TVaR_\alpha$, as defined in \eqref{proj2-TVaR-distortion}. Using \eqref{def_distortion_tvar} and \eqref{S_pi_1}, we obtain 
    \begin{equation*}
        \TVaR_\alpha(\Pi_{I}(X)) = \int_0^\infty g(S_{\Pi_{I}(X)}(x)) \d x = (1+\theta_1)a + \delta \int_{d_2}^{d_2 + u_I - d_I}g(S_X(x))\d x.
    \end{equation*}
    In addition, we have $\TVaR_\alpha(R_{I}(X)) = \int_0^{d_1}g(S_X(x))\d x + \int_{d_1+d_I}^{d_2}g(S_X(x))\d x.$
    Because of the comonotonic additivity of $\TVaR$, it then follows that
    \begin{equation}\label{TVaR_T}
    \begin{aligned}
        \TVaR_\alpha(T_{I}(X)) 
        = &\, \TVaR_\alpha(R_{I}(X) + \Pi_{I}(X))\\    
        = &\, \int_0^{d_1}g(S_X(x))\d x + \int_{d_1+d_I}^{d_2}g(S_X(x))\d x\\
          &\, + \delta \int_{d_2}^{d_2 + u_I - d_I}g(S_X(x))\d x + (1+\theta_1)a.
    \end{aligned}
    \end{equation}
    Thus, we have $\frac{\partial}{\partial d_1} \TVaR_\alpha(T_{I}(X)) = g(S_X(d_1)) - g(S_X(d_1 + d_I))$ and $\frac{\partial}{\partial d_2} \TVaR_\alpha(T_{I}(X)) = (1-\delta)g(S_X(d_2)) + \delta g(S_X(d_2+u_I-d_I))$.

    It then follows that
    \begin{equation}\label{dTVaR/dd1}
    \begin{aligned}
        \frac{\d}{\d d_1} \TVaR_\alpha(T_{I}(X)) 
        = &\, \frac{\partial}{\partial d_1} \TVaR_\alpha(T_{I}(X)) + \frac{\d d_2}{\d d_1}\frac{\partial}{\partial d_2} \TVaR_\alpha(T_{I}(X))\\
        = &\, [S_X(d_1) - S_X(d_1+d_I)]\left\{ \frac{g(S_X(d_1)) - g(S_X(d_1 + d_I))}{S_X(d_1) - S_X(d_1+d_I)}  \right.\\
        & \, \left. - \left[(1-\delta)\frac{g(S_X(d_2))}{S_X(d_2)} + \delta \frac{g(S_X(d_2+u_I-d_I))}{S_X(d_2)}\right] \right\}.
    \end{aligned}
    \end{equation}
    Let $v_\alpha \triangleq \VaR_\alpha(X)$. Depending on the value of $\alpha$, we have the following discussion. 
    \begin{enumerate}[(i)]
        \item If $d_1\leq v_\alpha-d_I$, we have $g(S_X(d_1)) = g(S_X(d_1 + d_I)) = 1 $, and thus $\frac{\d}{\d d_1} \TVaR_\alpha(T_{I}(X))\leq0$, which means $\TVaR_\alpha(T_{I}(X))$ decreases with $d_1$. 
        
        \item If $d_1\geq v_\alpha$, it then follows that $$\frac{\d}{\d d_1} \TVaR_\alpha(T_{I}(X)) = \frac{\delta}{\alpha}[S_X(d_1) - S_X(d_1+d_I)]\left\{ 1-\frac{S_X(d_2+u_I-d_I)}{S_X(d_2)} \right\}\geq0.$$
        
        \item If $v_\alpha-d_I \leq d_1 \leq v_\alpha$, we have 
        \begin{equation}\label{dtvar_dd1}
        \begin{aligned}
        \frac{\d}{\d d_1} \TVaR_\alpha(T_{I}(X)) &= \left[\frac{\alpha - S_X(d_1+d_I)}{S_X(d_1) - S_X(d_1+d_I)}-\delta\frac{S_X(d_2+u_I-d_I)}{S_X(d_2)}+\delta-1\right]\\
        &~~~~~~~~~~~~~~~~~~~~~~~~~\cdot\frac{1}{\alpha}[S_X(d_1) - S_X(d_1+d_I)].          
        \end{aligned}            
        \end{equation}
        Denote $$H(d_1)\triangleq \frac{\alpha - S_X(d_1+d_I)}{S_X(d_1) - S_X(d_1+d_I)}-\delta\frac{S_X(d_2+u_I-d_I)}{S_X(d_2)}+\delta-1.$$
        Assuming that $\frac{S_X(d+u_I-d_I)}{S_X(d)}$ is an increasing function of $d$, it follows from \eqref{dd2_dd1} that $H(d_1)$ is an increasing function of $d_1$. Furthermore, in Case (i), $H(d_1)\leq0$ when $d_1 = v_\alpha - d_I$ and in Case (ii), $H(d_1)\geq0$ when $d_1 = v_\alpha$. Hence, there exists $\bar{d_1}\in\mathbb R$ satisfying $H(\bar{d_1}) = 0$ such that $H(d_1) \leq 0$ when $d_1\leq \bar{d_1}$ and $H(d_1) \geq 0$ when $d_1\geq \bar{d_1}$. Therefore, from \eqref{dtvar_dd1}, we have that $\frac{\d}{\d d_1} \TVaR_\alpha(T_{I}(X))\leq0$ for $d_1\leq \bar{d_1}$ and $\frac{\d}{\d d_1} \TVaR_\alpha(T_{I}(X))\geq0$ for $d_1\geq \bar{d_1}$.
    \end{enumerate}
    Combining the above discussion, we have that $\TVaR_\alpha(T_{I}(X))$ decreases for $d_1\leq \bar{d_1}$ and increases for $d_1\geq \bar{d_1}$. If $\bar{d_1}<\tilde d$, the optimal parameters are $d_1^* = \bar{d_1}$ and $d_2^* = \bar{d_2}$, where $(\bar{d_1}, \bar{d_2})$ are determined by solving \eqref{proj2_op_parameters}. However, if $\bar{d_1}\geq\tilde d$, the optimal values are $d_1^* = \tilde d$ and $d_2^* = d_1^* + d_I$, yielding the optimal indemnity function $I^*(x) = (x-\tilde d)_+$. This completes the proof. 
\end{proof}

\begin{remark}

For a general distribution, the closed-forms of the optimal parameters are not available.
We propose an algorithm, which is designed based on equation \eqref{dTVaR/dd1}, to numerically calculate the values of those parameters. 
We use the ``bisection'' method to obtain the optimal indemnity function for a general risk distribution.
Note that the derivative of the objective function admits a unique root of $d_1$ by equation \eqref{dTVaR/dd1}. 
Then $d_2$ can be uniquely determined by $d_1$ through the equation $\mathbb E[I(X)] = a$ for a given $a$. The pseudocode of the process is shown in Algorithm 1.

\begin{algorithm}
\DontPrintSemicolon
\SetAlgoLined
\KwData{Parameters: $\mu$, $\Delta_a$, $\Delta_\delta$, $\theta_0$, $\theta_1$, $\theta_2$\;~~~~~~~~ Functions: g, S}
\KwResult{$a^*$, $d_1^*$, $d_2^*$}
\For{each $\delta$ in $[0, 1]$ with step $\Delta_\delta$}{
    \For{each $a$ in $[0, \mu]$ with step $\Delta_a$}{
        Solve for $\tilde{d}$ such that $\mathbb E[(X-\tilde d)_+] = a$\;
        $d_I \leftarrow \frac{\theta_1 - \theta_0 + \delta}{\delta} \cdot a$\;
        $u_I \leftarrow \frac{\theta_2 - \theta_0 + \delta}{\delta} \cdot a$\;
        \eIf{$d_I > 0$}{
            $d_1^*\leftarrow$Perform bisection on $d_1$ between $0$ and $\tilde{d}$ until $\frac{\d\rho_g(T_I(X))}{\d d_1} = 0$ in equation \eqref{dTVaR/dd1}. For each $d_1$, solve $d_2$ from $\int_{d_1}^{d_1+d_I}S(x)dx + \int_{d_2}^{\infty}S(x)dx = a$\;
        }{
            $d_I \leftarrow 0$\;
            $d_1^* \leftarrow \tilde{d}$\;
        }
        Compute $\rho_g(T_I(X))$ in equation \eqref{TVaR_T}\;
    }
    $a^*\leftarrow$ Select $a$ such that $\rho_g(T_I(X))$ reaches its minimum\;
    $d_1^*\leftarrow$ Select $d_1$ such that $\rho_g(T_I(X))$ reaches its minimum\;
    $d_2^*\leftarrow$ Solve $d_2$ such that $\int_{d_1^*}^{d_1^*+d_I}S(x)dx + \int_{d_2}^{\infty}S(x)dx = a^*$
}
\caption{Iterative Optimization of the Optimal Parameters of Indemnity Functions for a General Loss Distribution Given $\delta$}
\end{algorithm}

\end{remark}

\begin{example}
Assume that the loss follows an Exponential distribution, i.e., $X \sim \text{Exp}(1/\mu)$ with survival function $S_X(x) = e^{- x/\mu } $ for $x \geq 0 $, and $\mu = \E [ X ]$. 
It can be easily verified that $\frac{S_X(d+u_I-d_I)}{S_X(d)}$ is constant, which satisfies the condition in Proposition \ref{lemma_opt_parameters_1}. For a given $a\in[0,\mu]$, we would like to solve the problem \eqref{Proj2_prob_I1_a}. Noe that $\tilde{d} = \mu \ln \frac{\mu}{a}$ is obtained from $\mathbb{E}[(X - d)_+] = a$.
Solving the system of equations in \eqref{proj2_op_parameters}, we find the optimal parameters

\begin{align*}
\bar{d}_1 & = \mu \ln \left ( 1 - \delta + \delta e^{-\frac{d_I}{\mu }} + \delta e^{-\frac{u_I-d_I}{\mu}} - \delta e^{-\frac{u_I}{\mu}}\right ) - \mu\ln \alpha, \\
\bar{d}_2 & = - \mu \ln \left ( \frac{a}{\mu} - \frac{\alpha (1 - e^{-\frac{d_I}{\mu}}) }{1 - \delta + \delta e^{-\frac{d_I}{\mu }} + \delta e^{-\frac{u_I-d_I}{\mu}} - \delta e^{-\frac{u_I}{\mu}}}\right ).
\end{align*}
It is easy to see that  $\bar{d}_1$ and $\bar{d}_2$ are linearly increasing with respect to $\mu$ if the ratio $\frac{a}{\mu}= \frac{\E [I(X)]}{\E[X]}$ is fixed. 
This observation suggests that, given a fixed ratio $\frac{a}{\mu}$, the insurer would like to transfer large losses to the reinsurer, while retain relatively small losses. 
Figure \ref{Fig:I*_deltaFixed} illustrates the optimal indemnity function as defined in \eqref{I_opt_fixa_cvar} for $\delta = 1$. When $\bar d_1 < \tilde d$, the optimal indemnity follows a two-layer structure, while for $\bar d_1 \geq \tilde d$, it takes a form of a stop-loss function.

Furthermore, we use Algorithm 1 to numerically calculate the  optimal solution to the problem \eqref{Proj2_dedimension2}.
Under the setting in Figure \ref{Fig:I*_mu}, the optimal contract $ I^*(x) = (x - d_1^*)_+ - (x - d_1^* - d_I^*) _+ + (x - d_2^*)_+ $ is reduced to a stop-loss contract because $d_2^* =d_1^* + d_I^*$. From Figure \ref{Fig:I*_mu}-(a), we can see that  the value of $d_1^*$ increases linearly with respect to the mean of the exponential loss, and thus, the insurer transfers less loss to the reinsurer. 
Meanwhile, we calculate $a^* \triangleq \E [I^*(X) ]$ and present results in Figure \ref{Fig:I*_mu}-(b). It is interesting to see that the ratio $\frac{a^*}{ \mu }$ is a constant. This result suggests that it is always optimal for the insurer to transfer a certain percentage of the mean of this exponential loss. 

\begin{figure}[htbp]
    \centering
    \begin{subfigure}[b]{0.49\textwidth}
        \centering
        \includegraphics[width=\textwidth]{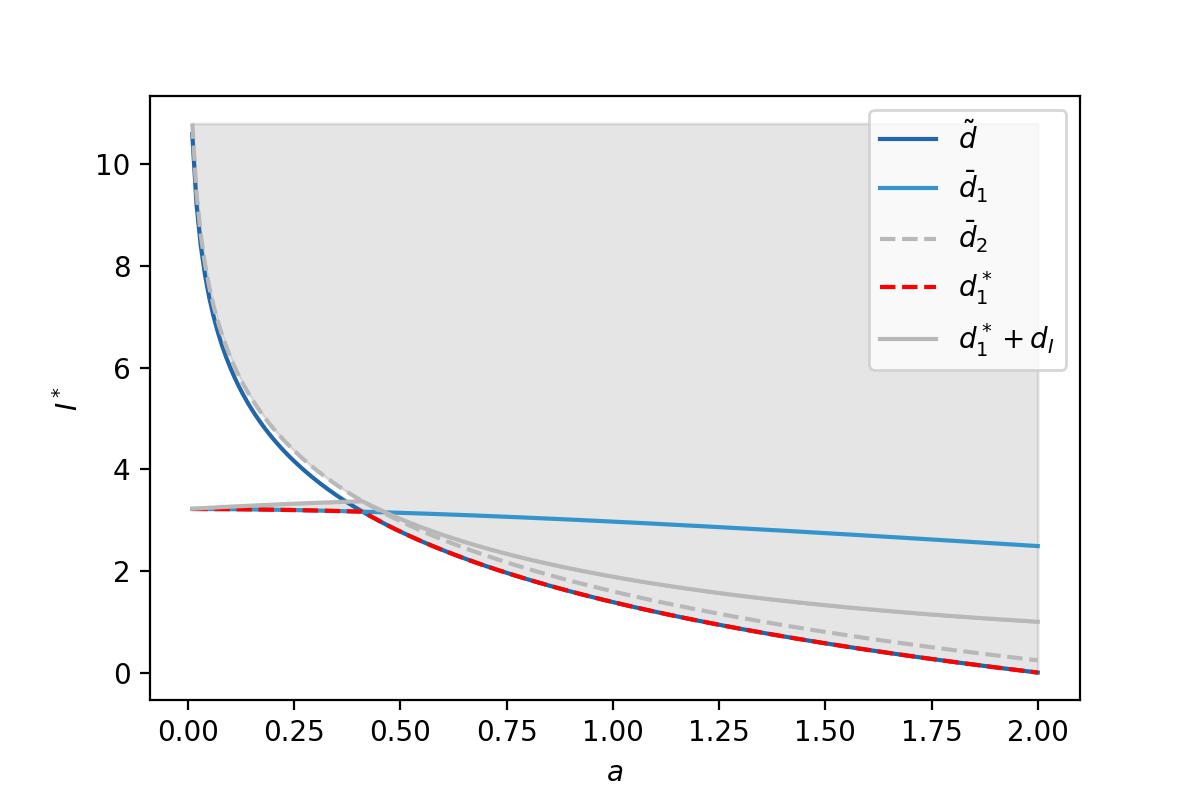} 
    \end{subfigure}
    \caption{The optimal indemnity function among all indemnities with the same expected value $a$, for the case when $\delta = 1$. The gray area represents the optimal ceded amount. The other parameters are set as follows: $\mu = 2$, $\theta_0 = 1$, $\theta_1 = 0.5$, $\theta_2 = 2$, and $\alpha = 0.2$.}
    \label{Fig:I*_deltaFixed}
\end{figure}

\begin{figure}[htbp]
    \centering
    \begin{subfigure}[b]{0.49\textwidth}
        \centering
        \includegraphics[width=\textwidth]{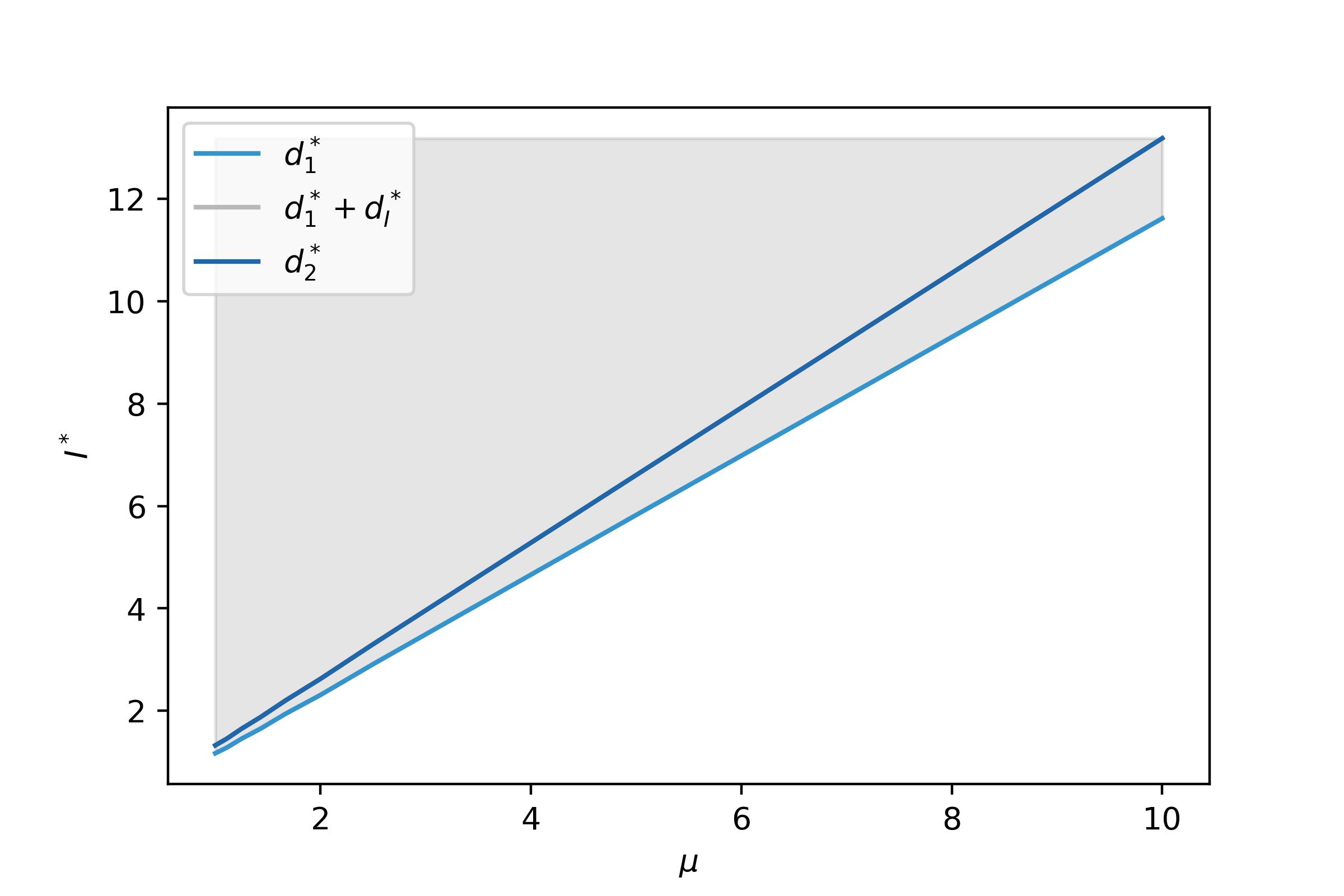} 
        \caption{}
    \end{subfigure}
    \hfill
    \begin{subfigure}[b]{0.49\textwidth}
        \centering
        \includegraphics[width=\textwidth]{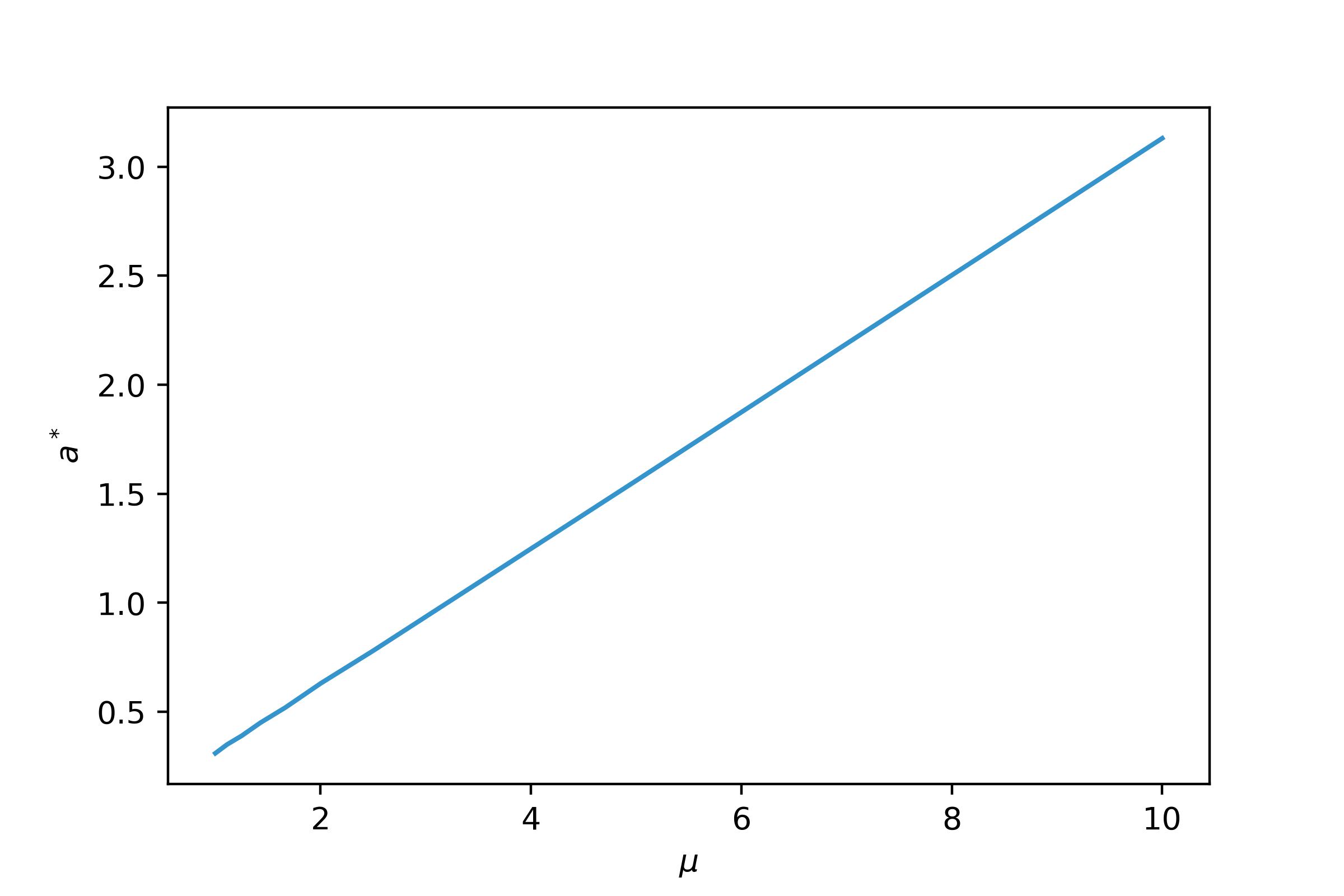} 
        \caption{}
    \end{subfigure}
    \caption{(a) The optimal indemnity function $I^*$ with respect to $\mu$. The gray area represents the optimal ceded amount. (b) The optimal expected ceded loss with respect to $\mu$. Here, $X\sim$Exp$(1/\mu)$. The parameters are set as follows: $\theta_0=1$, $\theta_1=0.5$, $\theta_2=2$, $\alpha_1=0.2$, $\delta=1$.}
    \label{Fig:I*_mu}
\end{figure}

\end{example}

\begin{example}
In \cite{CCT2016} the insurer is assumed to adopt the risk-adjusted liability as the optimization objective, and consider the problem
\begin{align}\label{risk-adjusted-obj}
\min_{I\in\mathcal I}\mathcal L(T_I(X)) = \min_{I\in\mathcal I}\{\mathbb E[T_I(X)] + \omega\cdot\rho(T_I(X) - \E[T_I(X)])\},
\end{align}
where $\rho$ is a risk measure, $\pi_{\text{RE}} (I(X))$ is the retrospective rating premium defined in \eqref{retro_CCT}, $ T_I(X) = X - I(X) + \pi_{\text{RE}} (I(X)) $ is the total risk exposure and $\omega\in(0,1)$ is known as the cost-of-capital rate.
When $\rho = \TVaR_\alpha$ and $X$ follows a Pareto distribution, the results of optimal solutions to \eqref{risk-adjusted-obj} are presented in Example 5.1 in \cite{CCT2016}. 
To facilitate a comparison between the frameworks used in \cite{CCT2016} and in our work, we follow the same distribution assumption of Example 5.1 in \cite{CCT2016}, and match parameters in our performance-based variable premium scheme with those used in retrospective rating premium \eqref{retro_CCT}. 
Then we numerically solve the problem
\begin{align}\label{ex3.2-opt prob}
\min_{I\in \mathcal{I}} \TVaR_\alpha (X - I(X) + \Pi_{I} (X ) ). 
\end{align}
The optimal solutions to problems \eqref{risk-adjusted-obj} and \eqref{ex3.2-opt prob} are both stop-loss contracts, denoted by $I_c^* (x) = (x-d_c^*)_+ $ and $ I^*(x) = (x - d^*)_+ $, respectively. 
 The comparison of $d_c^* $ and $d^*$  is presented in Table \ref{tab:compare results}, in which values of $d_c^* $ come from Example 5.1 in \cite{CCT2016}. 
Clearly, $d^* < d_c^*$ in all five cases,  which implies that the insurer in our model shall purchase more reinsurance coverage compared to the case in \cite{CCT2016}. 
 This result may be a consequence of that the problem \eqref{ex3.2-opt prob} aims to minimize  TVaR, which is a tail risk measure. Therefore, transferring large losses to the reinsurer can significantly reduce the insurer's right tail risk exposure. 
Such pattern is more obvious for large $\delta$, i.e., when the variable premium $\Pi_I(X)$ introduces more variation into the insurer's objective function.


\begin{table}
    \centering
    \begin{tabular}{c|c|cccc}\hline\hline
        $L = \delta/1.025$& $d^*$ &     $d^*_c$\\\hline
         0.1&  61.33&     66.54\\
         0.3&  40.33&   67.83\\
         0.5&  28.35&    67.71\\
         0.7&  21.77&     66.79\\
         0.9&  17.64&     66.03\\ \hline\hline
    \end{tabular}
    \caption{Comparison of the optimal deductibles for different scheme parameter $\delta$ under Pareto loss distribution.}
    \label{tab:compare results}
\end{table}
\end{example}

\section{Bowley-Optimal problem}\label{proj2-sec3}
As the seller, the reinsurer determines the parameters $\delta$ and $\theta_i$, for $i=0,1,2$, in the variable premium scheme \eqref{premium_I}. 
The parameter $\delta $ governs both the discount and penalty rates within the scheme. 
When the realized ceded loss $I(x) $ deviates from its expected value $\E [ I(X)]$, a larger $\delta$ amplifies the impact—resulting in either a greater penalty or a larger discount, depending on whether $\delta ( I(x) - \E [ I(X) ] ) $ is positive or negative. 
In other words, the reinsurer can adjust $\delta $ to control the premium charged in both the reward and the penalty scenarios. 
In the limiting case when $\delta = 0 $, the variable premium scheme simplifies to the expected-value premium principle  with risk loading $\theta_0$, resulting in a constant premium $\Pi_{I}(x)= \pi_0 (I(X)) = (1+\theta_0) \E [I(X)]$. 

In Section \ref{sec:insurer's opt prob}, we examine the insurer's optimization problem when the reinsurer implements this variable premium scheme.  
Obviously, the value of $\delta $ selected by the reinsurer influences the insurer's optimal solution, and, in turn, the indemnity function chosen by the insurer impacts the reinsurer's total risk exposure. 
A natural question arises: what benefit does the reinsurer gain from adopting this variable premium scheme, and what would be the optimal choice of $\delta $ from the reinsurer's perspective?
To explore this, we first modify Assumption \ref{assumption1} to allow flexibility in the choice of $\delta$.

\begin{assumption}\label{assumption2}
Assume that $ 0 \leq \bar{\theta}_1 \leq \theta_0 <\theta_2$ are given. 
For any $\delta \in [0, 1 ] $, take $\theta_1 = (\theta_0 - \delta ) \vee \bar{\theta}_1 $ and define
the corresponding  ``reward-and-penalty" variable premium scheme as follows. For any $I \in \cI$ and $ x \geq 0$, the premium scheme is given by
\begin{align}\label{premium_I-free delta}
    \Pi^\delta_I( x) & =  \min \left\{ \max \left\{ \pi_0 (I(X))+ \delta(I(x) - \mathbb E[I(X)]), \pi_1 (I(X)) \right \},  \pi_2 (I(X)) \right\} , 
\end{align}
where  $\pi_i (I(X)) = (1+\theta_i) \E [I(X) ]$ for $i=0,1,2$. 
\end{assumption}

We propose a Bowley optimization problem between the insurer and the reinsurer, structured as a two-step sequential process:
\begin{itemize}
\item[(i)] In the first phrase, the reinsurer chooses the value of $\delta \in [0,1] $ in the variable premium scheme under Assumption \ref{assumption2}. In response, the insurer selects the optimal indemnity function, denoted by $I^\delta $, which minimizes the following insurer's optimization problem
\begin{align}\label{BO-insurer prob}
        \min_{I\in \cI } \rho_1 (R_I(X) + \Pi^\delta_I  (X ) ),
\end{align}
where $\rho_1$ is the insurer's risk measure. 

\item[(ii)] In the second phrase, after observing the insurer’s choice $I^\delta$ as a function of $\delta$, the reinsurer then selects the optimal $\delta \in[0,1]$, which minimizes the following reinsurer's problem
\begin{align}\label{BO-reinsurer prob}
        \min_{ 0 \leq \delta \leq 1 } \rho_2 (I^\delta (X) - \Pi^\delta_{I^\delta }  (X) ),
\end{align}
where $\rho_2$ is the reinsurer's risk measure.
\end{itemize}

In the Bowley optimization problem, the reinsurer acts as the leader by setting the prices for reinsurance policies, while the insurer is the price taker. If there exists a value of $\delta$ and a reinsurance policy that allow both the insurer and reinsurer to optimize their respective objectives, as outlined in steps (i) and (ii),  then we say that this Bowley optimization problem has an optimal solution. 
Mathematically, we define Bowley-Optimality below.

\begin{definition}\label{Bowley-def}
A pair $ (\delta^*, I^* ) \in [0,1] \times \cI $ is said to be Bowley-Optimal (BO) if

\begin{itemize}
\item[(i)] $I^* = I^{\delta^*} $ is the optimal solution to the problem \eqref{BO-insurer prob} with respect to $\delta^*$, that is
$I^*  \in \arg\min_{I\in \cI } \rho_1 (R_I(X) + \Pi_I^{\delta^*}  (X) ) $. 

\item[(ii)] $\delta^* $ is the solution to the problem \eqref{BO-reinsurer prob}, that is
$\rho_2 (I^* (X) - \Pi^{\delta^*}_{I^*} (X) )  \leq \rho_2 \left (\tilde{I} (X) - \Pi^{\tilde{\delta} }_{\tilde{I}} (X) \right ) $ for all $ \tilde{\delta} \in [0,1]$ and $\tilde{I} = I^{\tilde{\delta}}\in  \arg\min_{I\in \cI } \rho_1\left (R_I(X) + \Pi^{\tilde{\delta}}_I(X) \right) $.
\end{itemize}
\end{definition}

The results from Section \ref{sec:insurer's opt prob} can be applied to further study the reinsurer's optimization problem in Definition \ref{Bowley-def}-(ii). 
Since the expected-value premium principle is a specific case of the variable premium scheme when $\delta = 0$, identifying the optimal solution $\delta^*$ in the Bowley-Optimal (BO) problem allows us to determine whether the variable premium scheme provides a greater benefit to the reinsurer compared to the expected-value premium principle.

Mathematically, the process to find the Bowley-Optimal (BO) solution begins by solving the problem \eqref{BO-insurer prob} to determine $I^\delta$ for each value of $\delta $. Next, this $I^\delta $ is substituted into the problem \eqref{BO-reinsurer prob} to find the optimal $\delta^* $. Finally, solving \eqref{BO-insurer prob} again using $\delta^* $ provides the optimal indemnity function $I^{\delta^*}$. 
Thus, the pair $(I^* = I^{\delta^*}, \delta^*) $ forms the BO solution. 
However, due to the complexity in the structure of $I^\delta$, determining $\delta^* $ analytically is challenging. Therefore, we rely on numerical examples to explore $\delta^* $ and the corresponding BO solution for both the insurer and reinsurer. 
In Examples \ref{example-PAR+TVaR}, \ref{example-exp+TVaR}, and \ref{example-EX+CVaR}, we adopt the following assumption for both parties:
\begin{align}\label{rho for examples}
    \rho_1 = \TVaR_\alpha , ~~~\text{ and }~~~\rho_2 = \TVaR_\beta , ~~~\text{ where } 0 < \alpha,\beta<1.
\end{align}
We then provide examples with both heavy-tailed and light-tailed distributions of $X$.

\begin{example}[Bowley-optimal problem under Pareto loss distribution]\label{example-PAR+TVaR}
    We assume that the loss $X$ follows the type II Pareto distribution with shape parameter $\zeta$ and scale parameter $\eta$, i.e. $X\sim$ Pa$(\zeta, \eta)$. The survival function of $X$ is given by $S_X(x) = \left(\frac{\eta}{x+\eta}\right)^\zeta$ for $x\geq 0$. Assume $\zeta>1$ to ensure that the expected value $\mathbb E[X]$ exists. 

    Figure \ref{Fig:I_opt} illustrates the retention levels in $I^\delta $, where $I^\delta \in  \arg\min_{I\in \cI } \TVaR_\alpha (R_I(X) + \Pi_I^{\delta}  (X ) )$ is the optimal solution for the insurer based on different values of $\delta \in [0,1]$.
    A higher $\delta$ leads to a stronger power of discount and penalty. 
    It is evident that as $\delta $ increases, the insurer prefers to retain more risks. In other words, $I^{\delta_1 } \geq I^{\delta_2} $ for $\delta_1 \leq \delta_2$. 
    In the scenario depicted in Figure \ref{Fig:I_opt}-(a), the optimal solutions are stop-loss functions for all values of $\delta$. 
    In contrast, in Figure \ref{Fig:I_opt}-(b), $I^\delta$ begins as a stop-loss function for small values of $\delta$ and transitions into a two-layer function as $\delta$ increases. 
    
    As illustrated in Figure \ref{Fig:rhoin_opt}, the value function of the insurer $\TVaR_\alpha(R_{I^\delta} (X) + \Pi^\delta_{I^\delta} (X))$ increases with $\delta$. 
    Since the optimal amount ceded decreases as $\delta$ rises, the insurer bears more risks from retained loss while facing reduced risk from premium payments. 
    Figure \ref{Fig:rhoin_opt} indicates that the increased risk from the retained loss outweighs the change in the risk of the premium. 
    Therefore, the insurer favors a constant premium scheme over a variable premium scheme.

    Conversely, the objective of the reinsurer $\TVaR_\beta(I^\delta (X) - \Pi^\delta_{I^\delta} (X))$ significantly benefits from adopting the variable premium scheme, as illustrated in Figure \ref{Fig:rhore_opt}. In Figure \ref{Fig:rhore_opt}-(a), the reinsurer's objective reaches its minimum at $\delta^* = 0.259$, the unique solution for this parameter set. In Figure \ref{Fig:rhore_opt}-(b), however, $\delta^*$ is not unique. Similar to the impact of $\delta$ on the insurer's value function, the reduction in $\TVaR_\beta(I^\delta(X))$ outweighs the increase in  $\TVaR_\beta(\Pi^\delta_{I^\delta} (X))$ as $\delta$ increases, resulting in an overall decrease in the reinsurer's objective. 
    In the context of the Bowley optimum, there may exist multiple optimal $\delta^*$ that minimize the reinsurer's objective. However, since the insurer's value function rises with increasing $\delta$, the smallest value within the optimal set of $\delta^*$ is preferred, as it is more likely to align the interests of both the insurer and reinsurer, facilitating agreement on the reinsurance contract.
    
    When the insurer's risk preference, represented by $\alpha$, remains constant, the variation in the optimal $\delta^*$ as the reinsurer's risk preference, represented by $\beta$, changes is shown in Figure \ref{Fig:delta_beta}. As $\beta$ increases, the minimum optimal $\delta^*$ initially rises and then falls. When $\beta = 1$, in which case $\TVaR_1 = \E $,  the reinsurer becomes risk neutral and prefers a constant premium scheme. This result suggests that the variable premium scheme helps the reinsurer mitigate risk whenever she is not risk neutral.

\begin{figure}[htbp]
    \centering
    \begin{subfigure}[b]{0.49\textwidth}
        \centering
        \includegraphics[width=\textwidth]{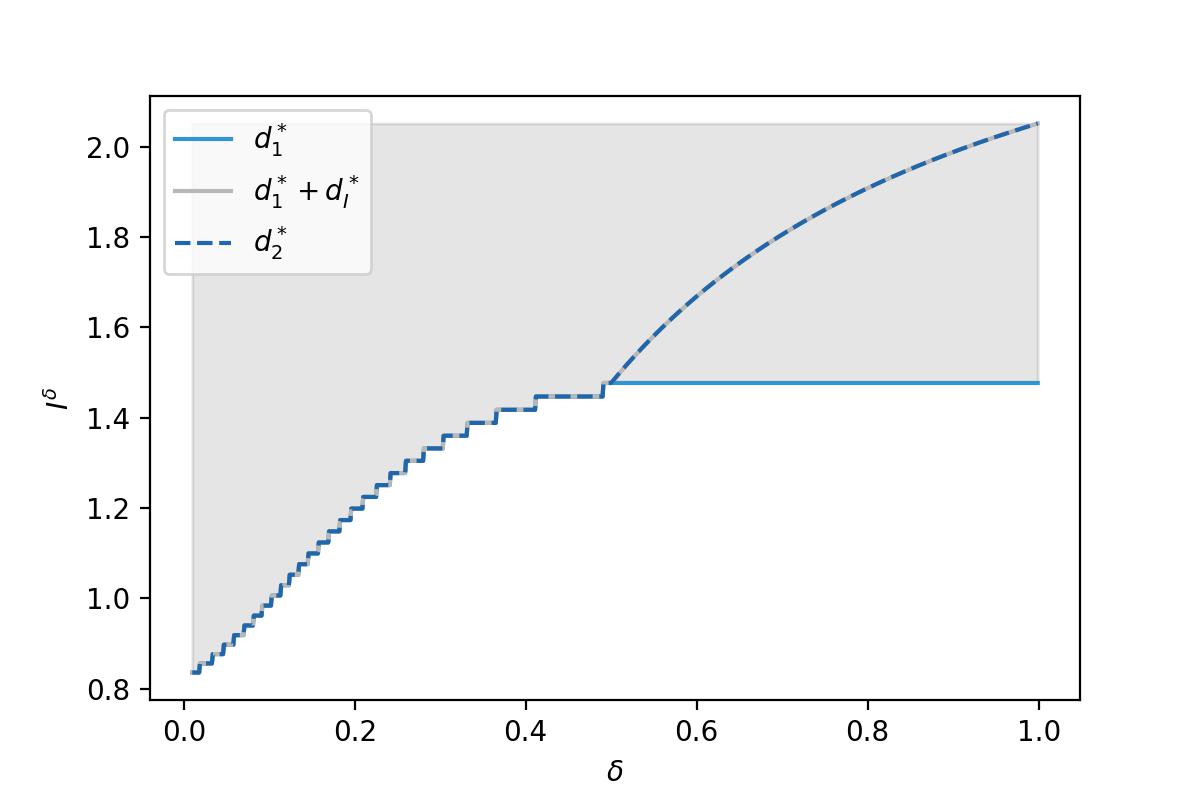} 
        \caption{}
        \label{fig:I_a}
    \end{subfigure}
    \hfill
    \begin{subfigure}[b]{0.49\textwidth}
        \centering
        \includegraphics[width=\textwidth]{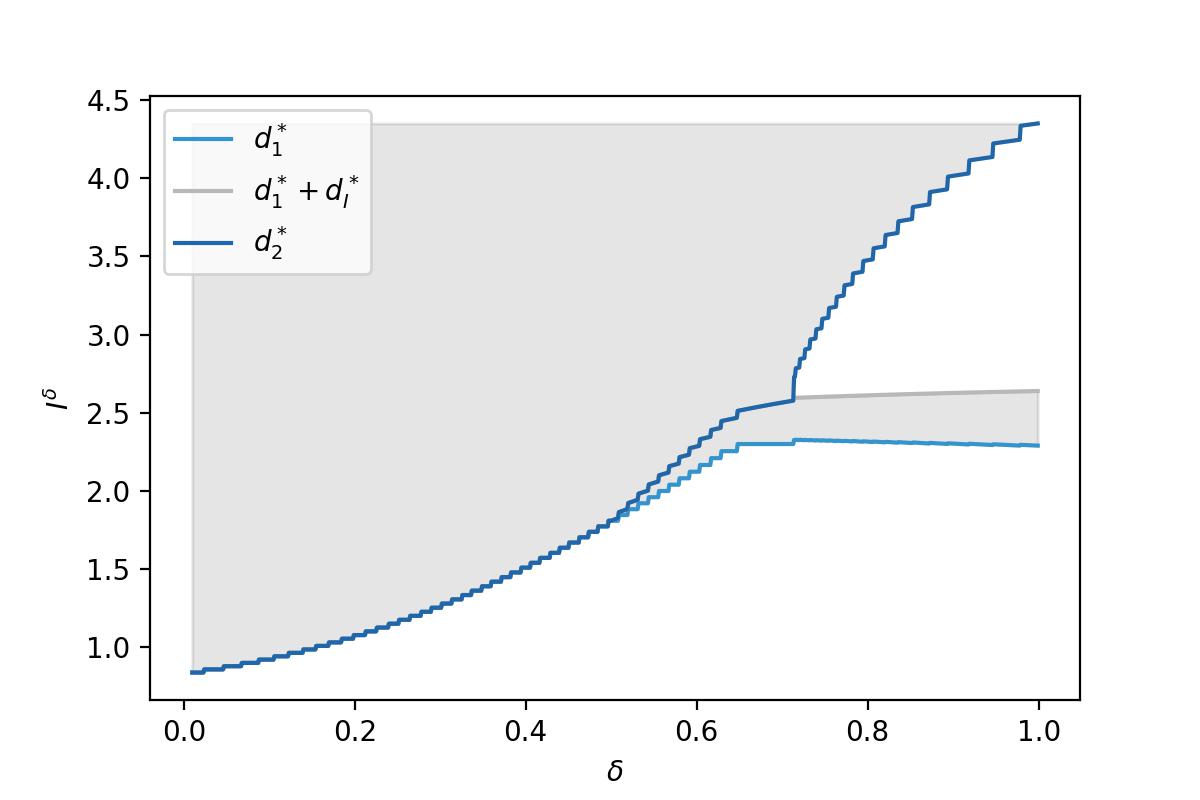} 
        \caption{}
        \label{fig:I_b}
    \end{subfigure}
    \caption{The optimal ceded loss function $I^\delta(x) = (x-d_1^*)_+ - (x - d_1^* - d_I^*)_+ + (x - d_2^*)_+$ with respect to $\delta$, where $d_I^* = \frac{\theta_1-\theta_0+\delta}{\delta}\mathbb E[I^\delta(X)]$. The shaded area represents the ceded portion. Here, $\rho = \TVaR_\alpha$ is adopted to measure the insurer's risk and $X\sim\text{Pa}(\eta, \zeta)$. The parameters are as follows. (a) $\eta=2$, $\zeta=2$, $\theta_0=1$, $\theta_1=0.5$, $\theta_2=2$, $\alpha=0.1$; (b) $\eta=2$, $\zeta=2$, $\theta_0=1$, $\theta_1=0.5$, $\theta_2=5$, $\alpha=0.2$.}
    \label{Fig:I_opt}
\end{figure}

\begin{figure}[htbp]
    \centering
    \begin{subfigure}[b]{0.49\textwidth}
        \centering
        \includegraphics[width=\textwidth]{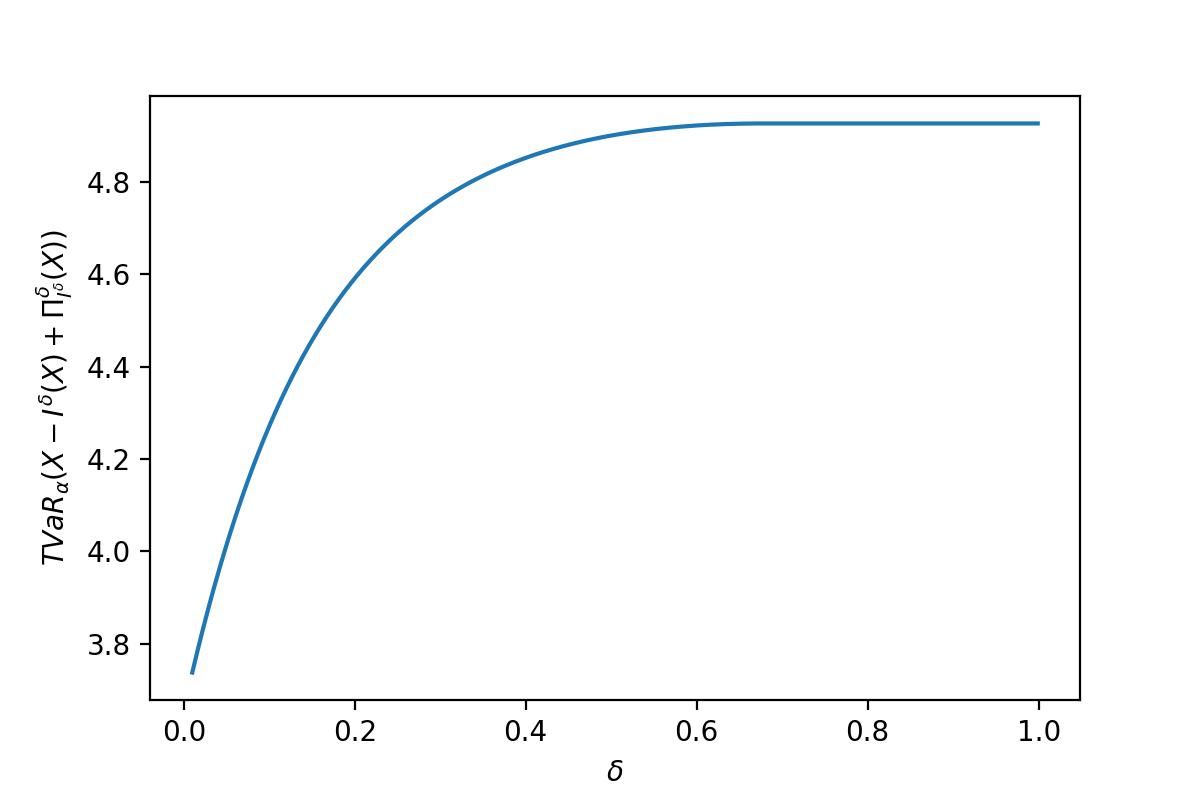} 
        \caption{}
        \label{fig:rhoin_a}
    \end{subfigure}
    \hfill
    \begin{subfigure}[b]{0.49\textwidth}
        \centering
        \includegraphics[width=\textwidth]{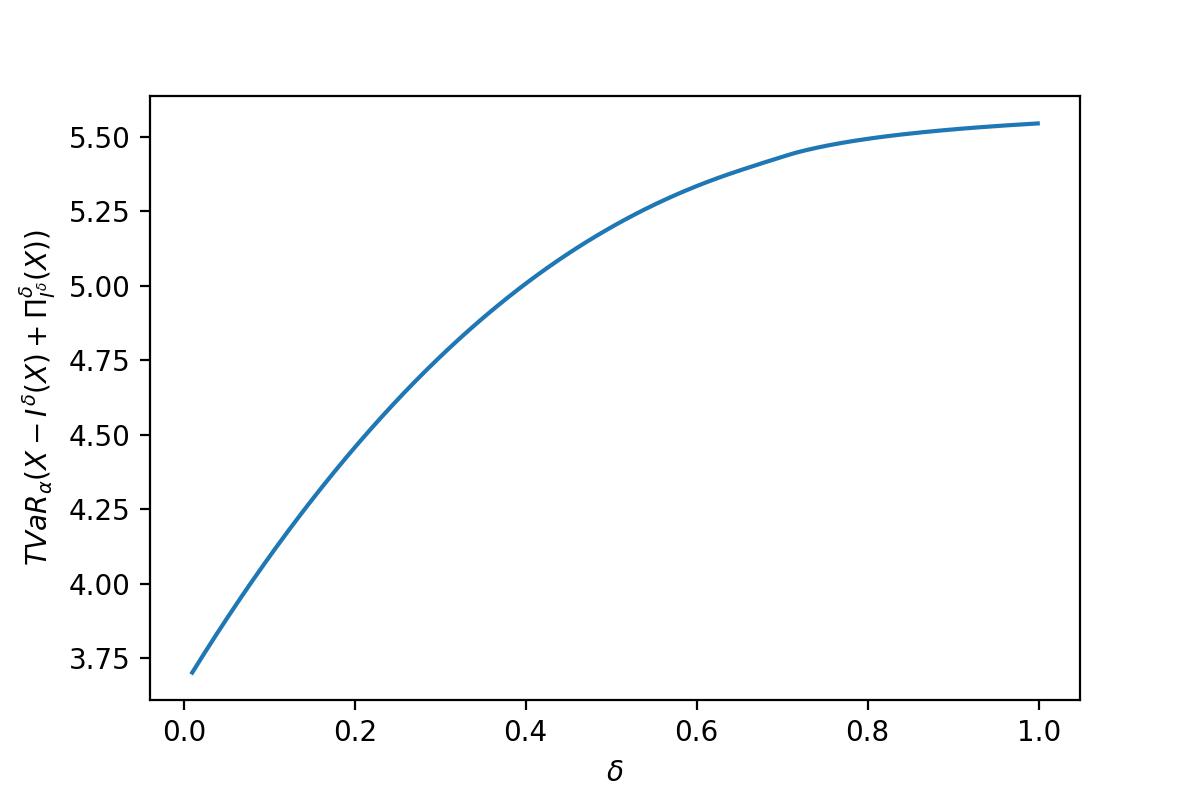} 
        \caption{}
        \label{fig:rhoin_b}
    \end{subfigure}
    \caption{The value function of the insurer with respect to $\delta$. Here, $\rho = \TVaR_\alpha$ is adopted by the insurer and $X\sim\text{Pa}(\eta, \zeta)$. The parameters are as follows. (a) $\eta=2$, $\zeta=2$, $\theta_0=1$, $\theta_1=0.5$, $\theta_2=2$, $\alpha=0.1$; (b) $\eta=2$, $\zeta=2$, $\theta_0=1$, $\theta_1=0.5$, $\theta_2=5$, $\alpha=0.2$.}
    \label{Fig:rhoin_opt}
\end{figure}

\begin{figure}[htbp]
    \centering
    \begin{subfigure}[b]{0.49\textwidth}
        \centering
        \includegraphics[width=\textwidth]{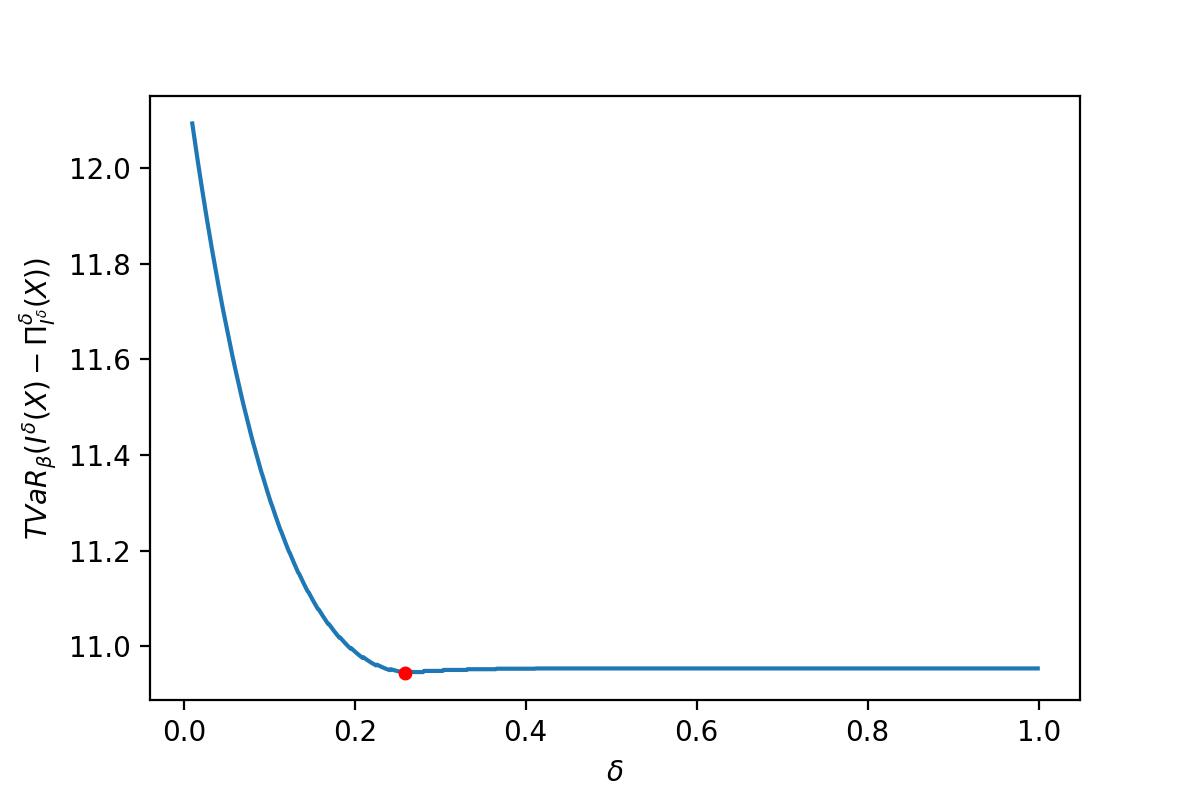} 
        \caption{}
        \label{fig:rhore_a}
    \end{subfigure}
    \hfill
    \begin{subfigure}[b]{0.49\textwidth}
        \centering
        \includegraphics[width=\textwidth]{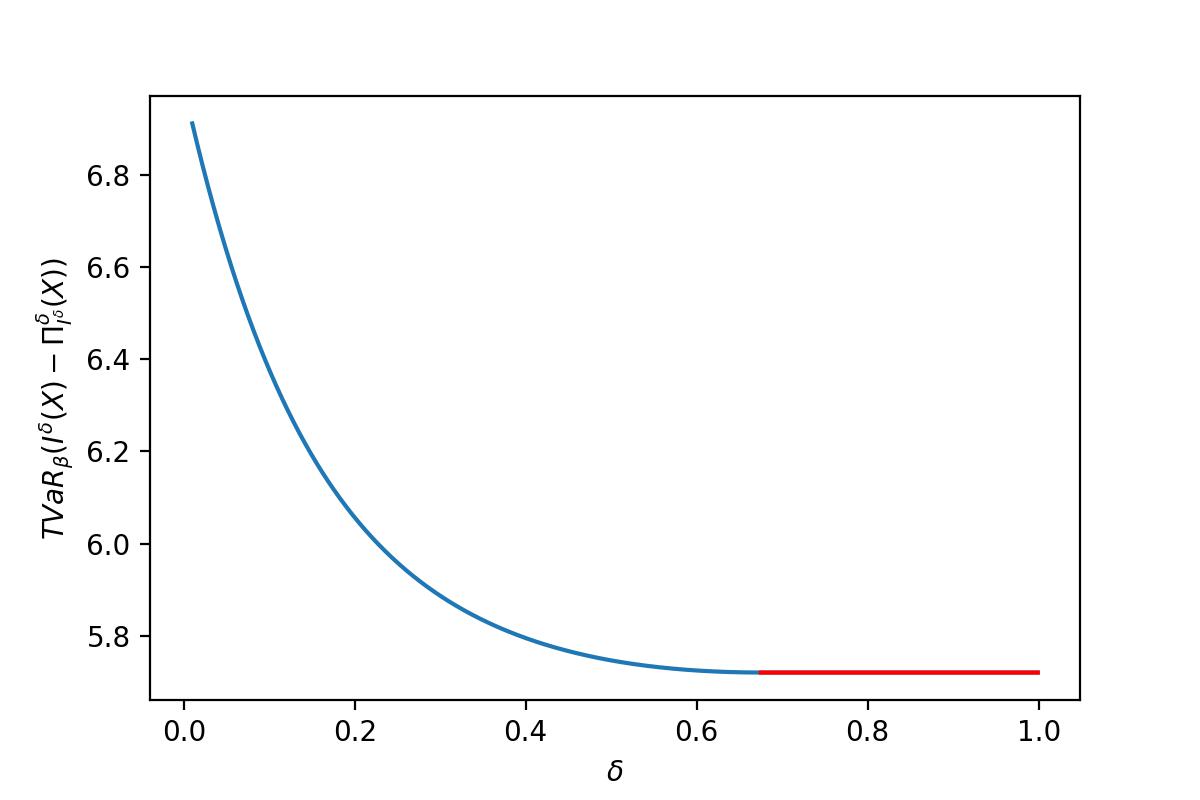} 
        \caption{}
        \label{fig:rhore_b}
    \end{subfigure}
    \caption{Assume $X\sim\text{Pa}(\eta, \zeta)$. The insurer adopts $\rho = \TVaR_\alpha$ and the reinsurer uses $\rho = \TVaR_\beta$ to measure their risk. The indemnity function $I^\delta$ minimizes the insurer's objective for a given $\delta$. The figure illustrates reinsurer's objective function with respect to $\delta$ when adopting $I^\delta$. The red represents the minimum value. The parameters are as follows. (a) $\eta=2$, $\zeta=2$, $\theta_0=1$, $\theta_1=0.5$, $\theta_2=2$, $\alpha=0.1$, $\beta = 0.05$; (b) $\eta=2$, $\zeta=2$, $\theta_0=1$, $\theta_1=0.5$, $\theta_2=2$, $\alpha=0.1$, $\beta = 0.1$.}
    \label{Fig:rhore_opt}
\end{figure}

\begin{figure}[htbp]
    \centering
        \centering
        \includegraphics[width=0.49\textwidth]{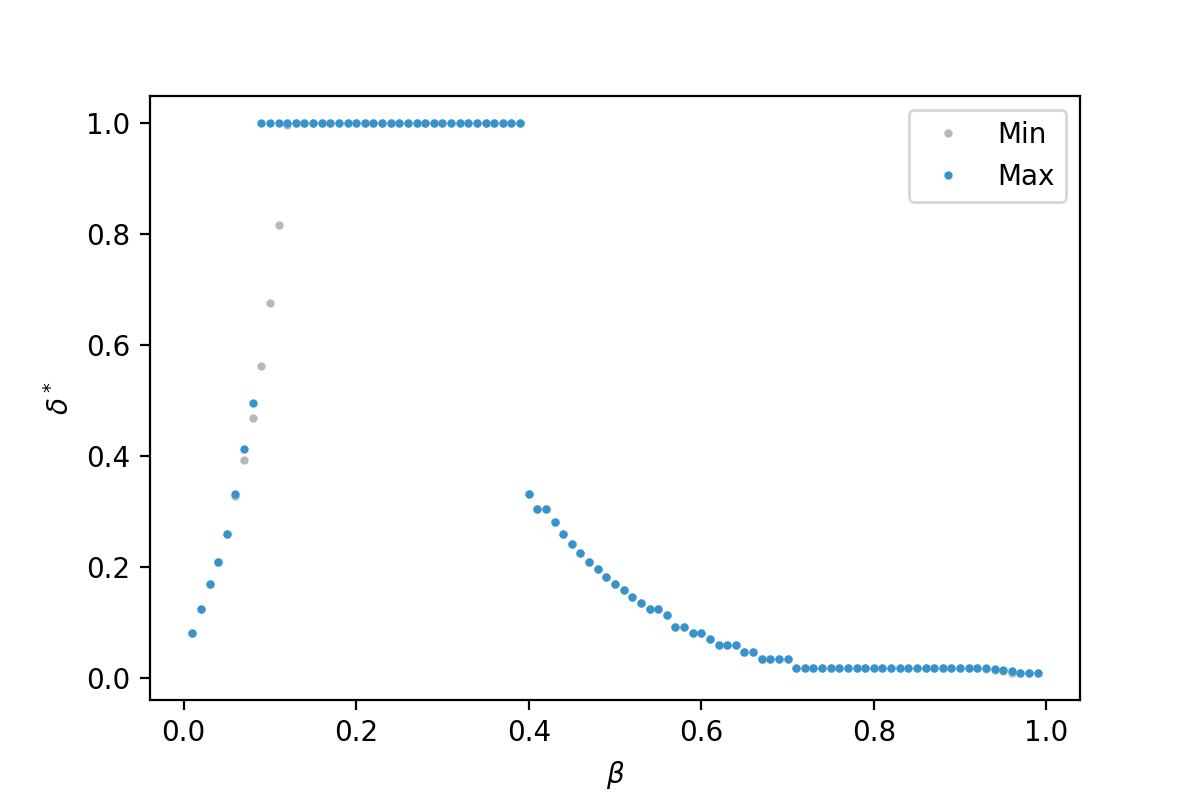} 
    \caption{The optimal $\delta^*$ with respect to $\beta$ under given $\alpha$. Here, assume $X\sim\text{Pa}(\eta, \zeta)$ and the insurer and the reinsurer adopt $\rho = \TVaR$ with levels $\alpha$ and $\beta$, respectively. The grey dots represent the minimum optimal $\delta^*$, while the blue dots represent the maximum. The dots in between indicate the other optimal values. The parameters are as follows. $\eta=2$, $\zeta=2$, $\theta_0=1$, $\theta_1=0.5$, $\theta_2=2$, $\alpha=0.1$.}
    \label{Fig:delta_beta}
\end{figure}
\end{example}

\begin{example}[Bowley-optimal problem under Exponential loss distribution]\label{example-exp+TVaR}
Assume  that \eqref{rho for examples} holds and 
$X \sim \text{Exp}(\lambda)$ with $S_X(x) = e^{-\lambda x } $ for $x \geq 0 $. 

As illustrated in Figure \ref{Fig:I_opt_exp}-(a), the optimal indemnity function is a stop-loss function for all $\delta$. 
In Figure \ref{Fig:I_opt_exp}-(b), with a larger value of $\theta_2$, the function $I^\delta $ becomes a two-layer function when $\delta $ is large. These findings are consistent with those from the Pareto loss distribution in Example \ref{example-PAR+TVaR}. However, unlike the Pareto distribution with the same expected value, the insurer tends to retain more loss under the Exponential loss distribution. This is intuitive, as the Exponential distribution has a lighter tail compared to the Pareto distribution, implying a lower risk of extreme losses.

With the same expectation and consistent conditions, the optimal $\delta^*$ for the Exponential loss distribution is notably smaller than that for the Pareto distribution, as seen in the comparison between Figures \ref{Fig:delta_beta} and \ref{Fig:delta_beta_exp}. This is likely because the insurer prefers to retain more risk under the Exponential distribution. A smaller $\delta^*$ may encourage the insurer to cede more loss, enabling the reinsurer to increase income by charging a higher premium.

\begin{figure}[htbp]
    \centering
    \begin{subfigure}[b]{0.49\textwidth}
        \centering
        \includegraphics[width=\textwidth]{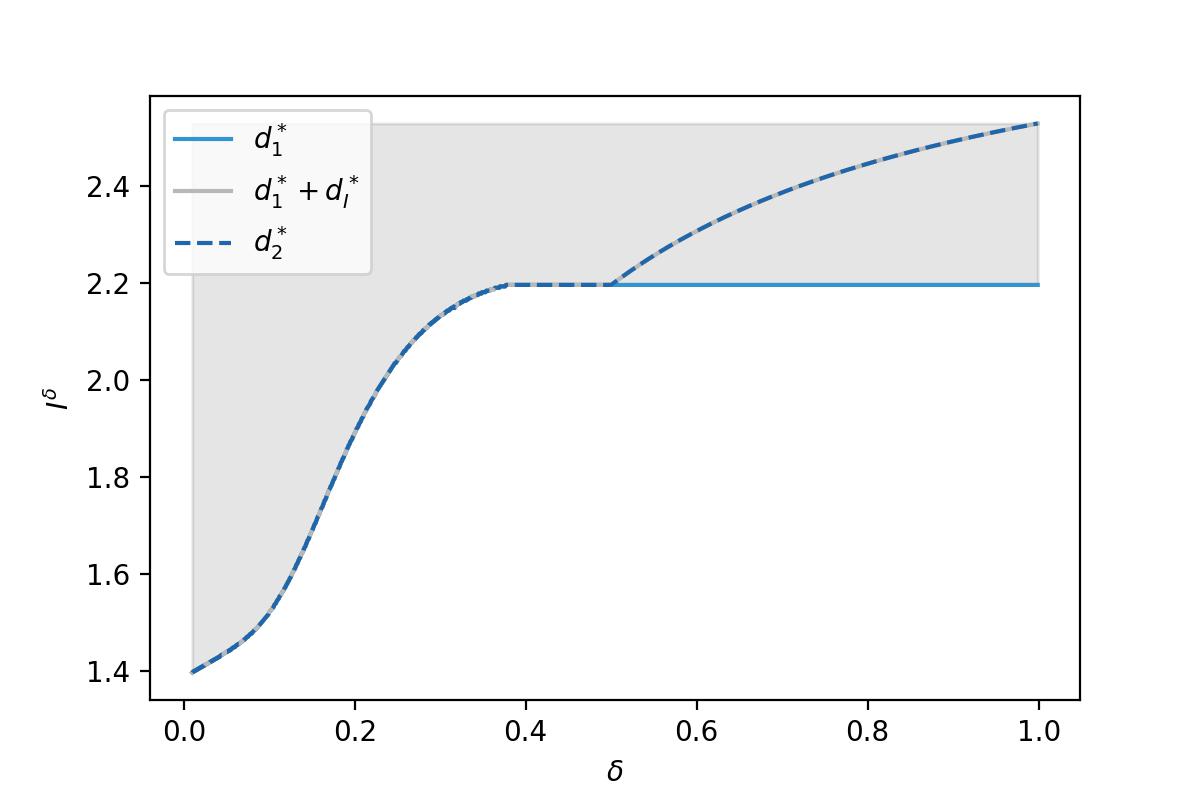} 
        \caption{}
        \label{fig:I_opt_exp_a}
    \end{subfigure}
    \hfill
    \begin{subfigure}[b]{0.49\textwidth}
        \centering
        \includegraphics[width=\textwidth]{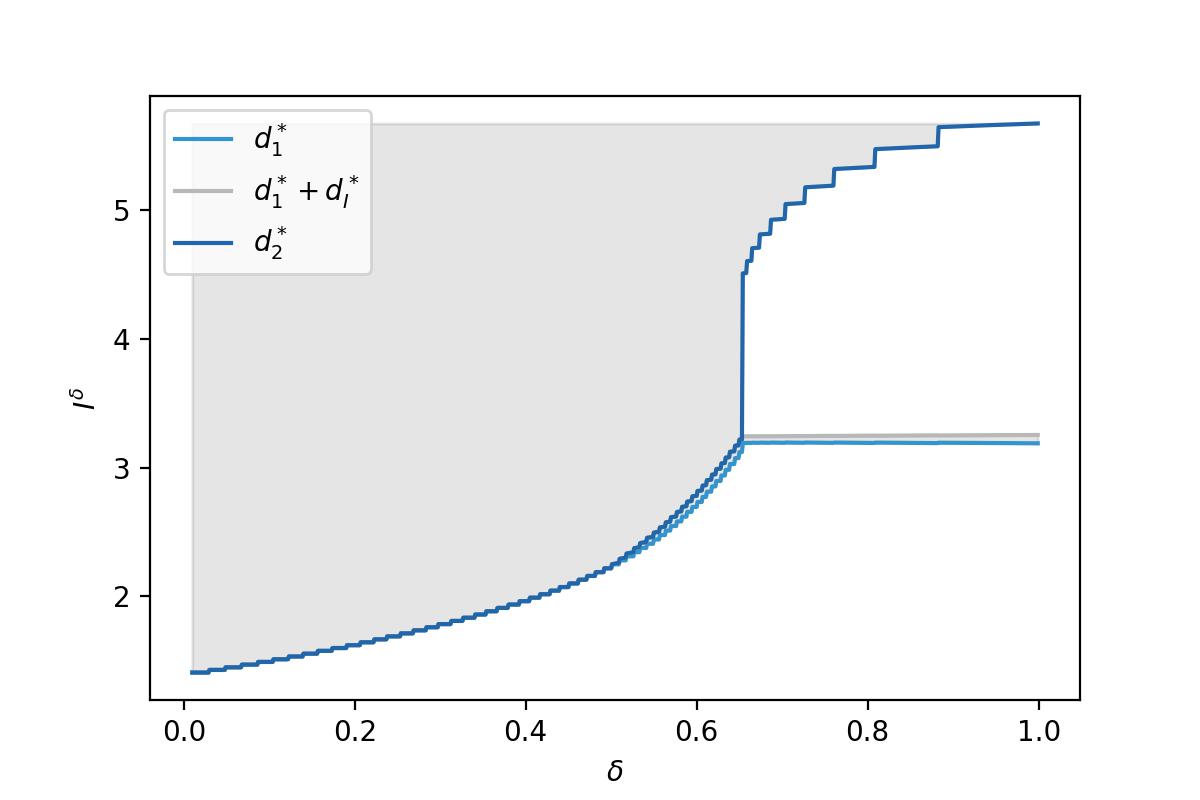} 
        \caption{}
        \label{fig:I_opt_exp_b}
    \end{subfigure}
    \caption{The optimal ceded loss function $I^\delta(x) = (x-d_1^*)_+ - (x - d_1^* - d_I^*)_+ + (x - d_2^*)_+$ with respect to $\delta$, where $d_I^* = \frac{\theta_1-\theta_0+\delta}{\delta}\mathbb E[I^\delta(X)]$. The shaded area represents the ceded portion. Here, $X\sim \text{Exp}(\lambda)$ and the insurer adopts $\rho = \TVaR_\alpha$. The parameters are as follows. (a) $\lambda = 0.5$, $\theta_0=1$, $\theta_1=0.5$, $\theta_2=2$, $\alpha=0.1$; (b) $\lambda = 0.5$, $\theta_0=1$, $\theta_1=0.5$, $\theta_2=10$, $\alpha=0.2$.}
    \label{Fig:I_opt_exp}
\end{figure}

\begin{figure}[htbp]
    \centering
        \centering
        \includegraphics[width=0.49\textwidth]{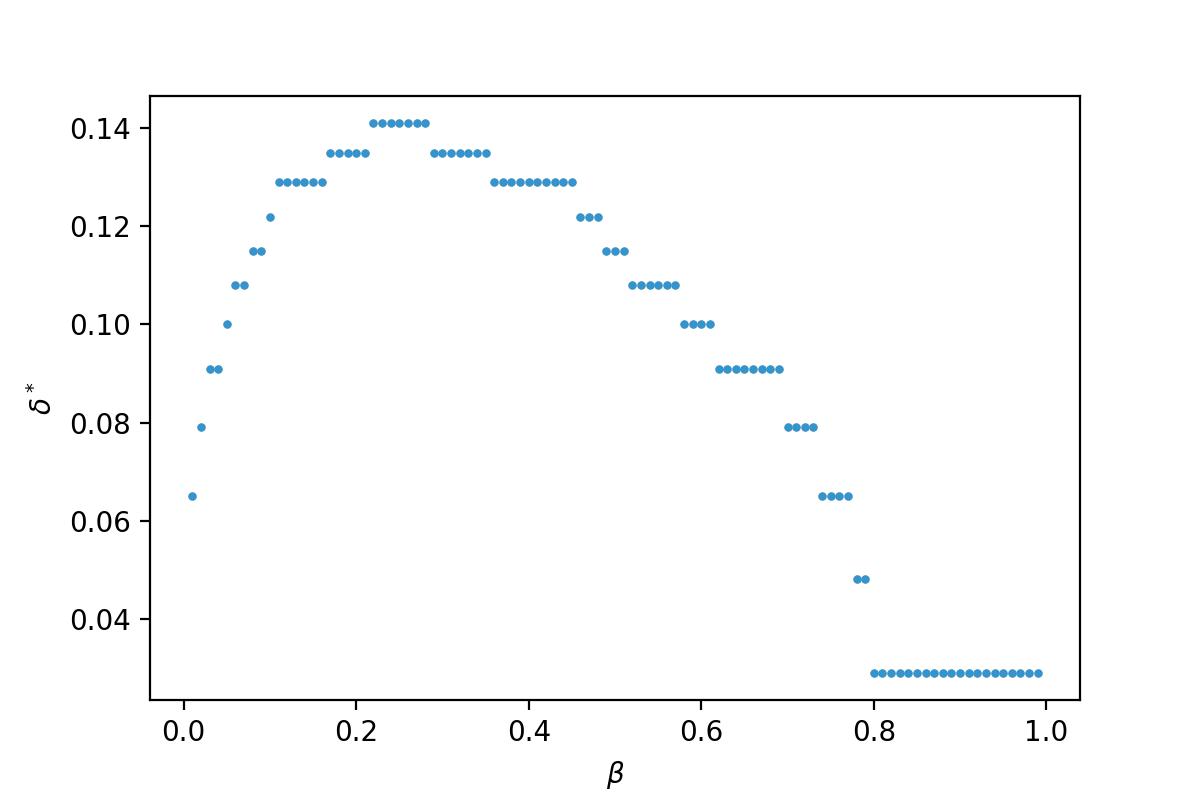} 
    \caption{The optimal $\delta^*$ that minimizes the reinsurer's objective function with respect to $\beta$. Here, assume $X\sim\text{Exp}(\lambda)$. The insurer and the reinsurer use $\TVaR$ to measure their risk with levels $\alpha$ and $\beta$, respectively.  The parameters are as follows. $\lambda=0.5$, $\theta_0=1$, $\theta_1=0.5$, $\theta_2=2$, $\alpha=0.1$.}
    \label{Fig:delta_beta_exp}
\end{figure}
\end{example}

\begin{figure}[htbp]
    \centering
    \begin{subfigure}[b]{0.49\textwidth}
        \centering
        \includegraphics[width=\textwidth]{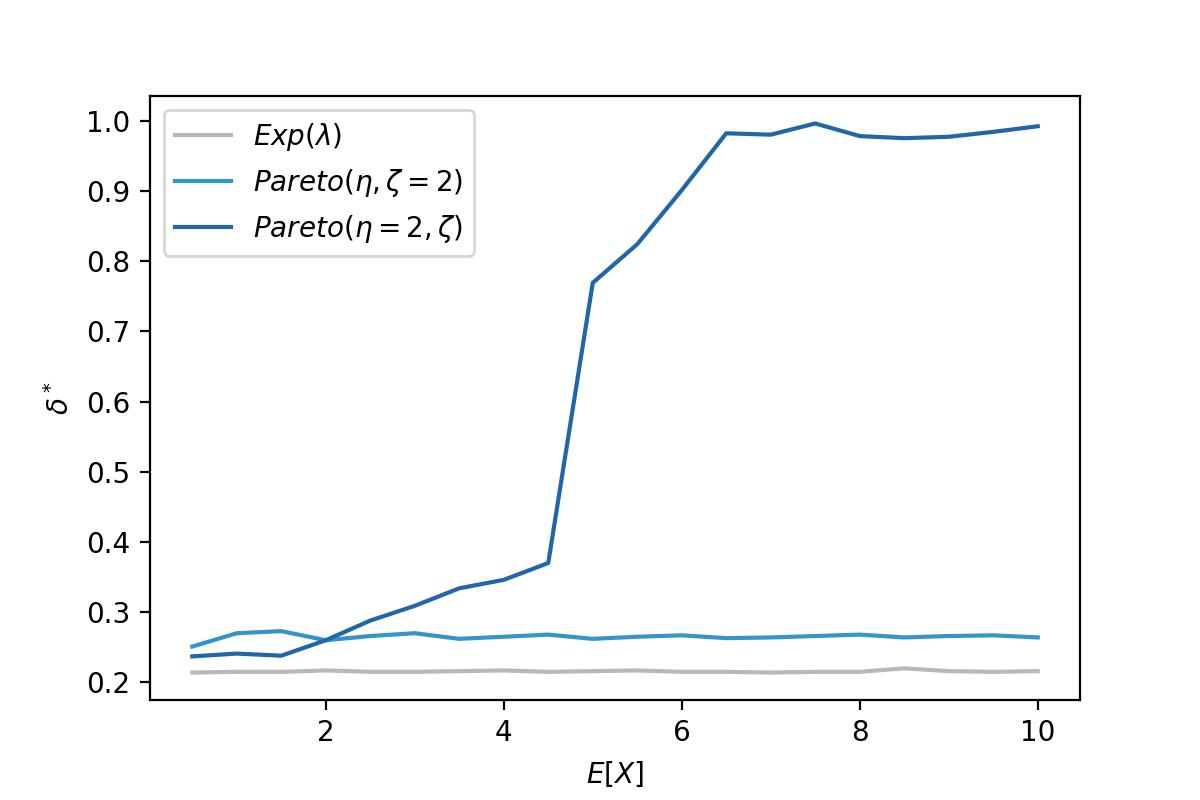} 
        \caption{}
        \label{fig:delta_a_EX_a}
    \end{subfigure}
    \hfill
    \begin{subfigure}[b]{0.49\textwidth}
        \centering
        \includegraphics[width=\textwidth]{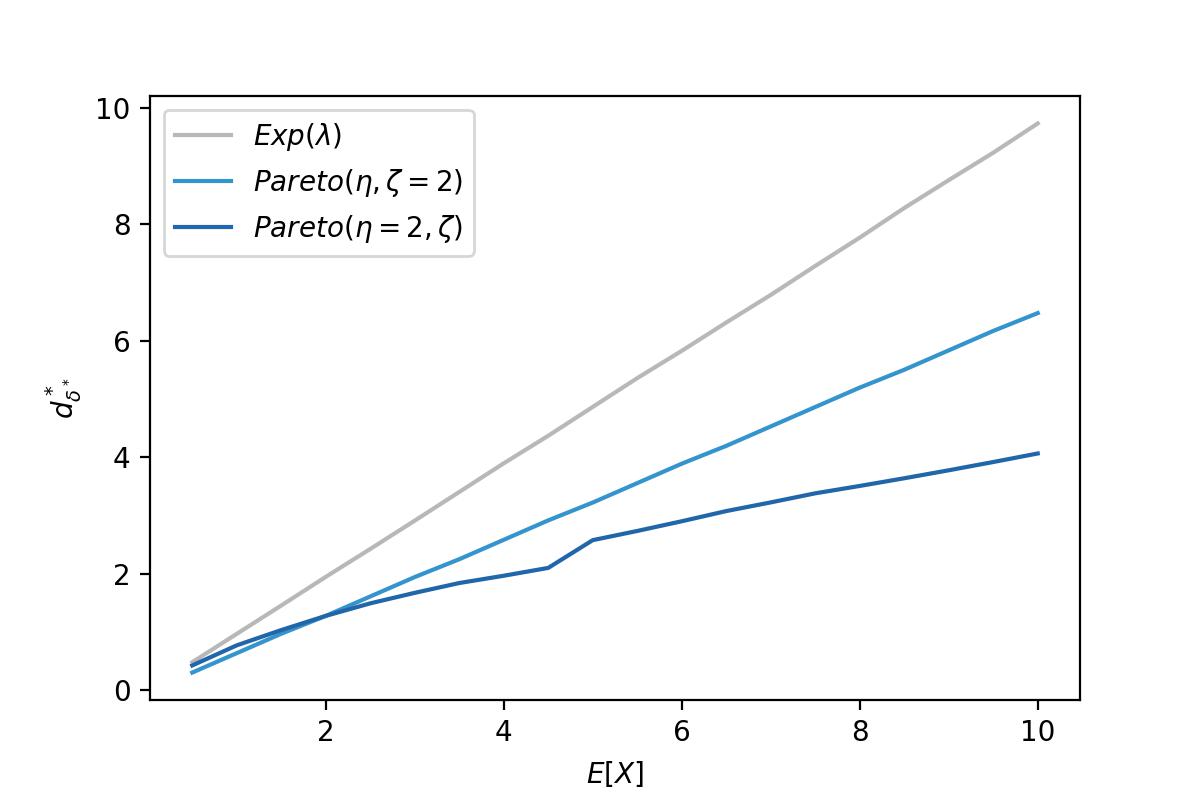} 
        \caption{}
        \label{fig:delta_a_EX_b}
    \end{subfigure}
    \caption{(a) The optimal $\delta^*$ with respect to $\mathbb E[X]$; (b) The optimal deductible $d^*_{\delta^*}$ with respect to $\mathbb E[X]$. If $X\sim\text{Exp}(\lambda)$, the results are illustrated by the grey line. If $X \sim \text{Pa}(\eta, \zeta)$, given $\mathbb E[X]$, the results of fixing $\zeta$ or $\eta$ while varying the other parameter are represented by the light blue and dark blue lines, respectively. The parameters are as follows. $\theta_0=1$, $\theta_1=0.5$, $\theta_2=2$, $\alpha=0.1$, $\beta = 0.05$.}
    \label{fig:delta_a_EX}
\end{figure}

\begin{example}[Other discussions on the Bowley-optimum]\label{example-EX+CVaR}
    Following Examples \ref{example-PAR+TVaR} and \ref{example-exp+TVaR}, we now discuss the impact of $\mathbb E[X]$ on the BO solution. We take the Pareto and Exponential distributions to illustrate our findings. 
    
    The Pareto distribution is characterized by two parameters, and we vary one while keeping the other constant. Figure \ref{fig:delta_a_EX}-(a) displays the variation of the optimal $\delta^*$, which achieves the Bowley optimum, in relation to changes in $\mathbb{E}[X]$. Under these parameter settings, the optimal indemnity functions at the Bowley optimum are all stop-loss functions. The deductible of these stop-loss functions, as influenced by $\mathbb{E}[X]$, is shown in Figure \ref{fig:delta_a_EX}-(b).
    
    As shown in Figure \ref{fig:delta_a_EX}, when the shape parameter $\zeta$ of the Pareto distribution is varied while keeping the scale parameter $\eta$ fixed, the optimal $\delta^*$ and the optimal deductible $d^*_{\delta^*}$ change significantly with shifts in $\mathbb{E}[X]$. Specifically, as the loss distribution becomes heavier-tailed (i.e., when $\zeta$ decreases as $\mathbb{E}[X]$ increases), the insurer decides to cede a larger proportion of the loss to the reinsurer, while the reinsurer demands a significantly higher $\delta^*$. In contrast, varying the scale parameter of either the Exponential or Pareto distribution only slightly affects the optimal $\delta^*$. The optimal deductible $d^*_{\delta^*}$ appears proportional to $\mathbb{E}[X]$ when only the scale parameter is altered, but this proportionality breaks down when the shape parameter is varied.
\end{example}

\begin{example}[Bowley-optimum with power distortion risk measure]
In this example, instead of using TVaR, we assume that both the insurer and reinsurer adopt power distortion risk measures. Specifically, the insurer's risk measure is defined as $\rho_1 (X) =\int_0^\infty [S_X(x)]^\alpha\d x $
and the reinsurer's risk measure is $\rho_2 (X) =\int_0^\infty [S_X(x)]^\beta\d x $, where $0<\alpha,\beta<1$ represent the risk aversion parameters of the insurer and reinsurer, respectively.
The underlying loss is assumed to follow an exponential distribution with scale parameter $\lambda$, that is, $ X \sim \text{Exp} (\lambda) $. 

All results are consistent with those obtained using $\TVaR$, with one key difference: the optimal $\delta^*$ for the Bowley-optimal problem decreases as $\beta$ increases when the power distortion risk measure is used. This is illustrated in Figure \ref{Fig:delta_beta_exp_power}. As the reinsurer approaches risk neutrality, a variable premium becomes unnecessary to optimize her objective. However, if the reinsurer is highly risk-averse (i.e., when $\beta$ is small), the variable premium scheme becomes significantly valuable in managing risk exposure.

\begin{figure}[htbp]
    \centering
        \centering
        \includegraphics[width=0.49\textwidth]{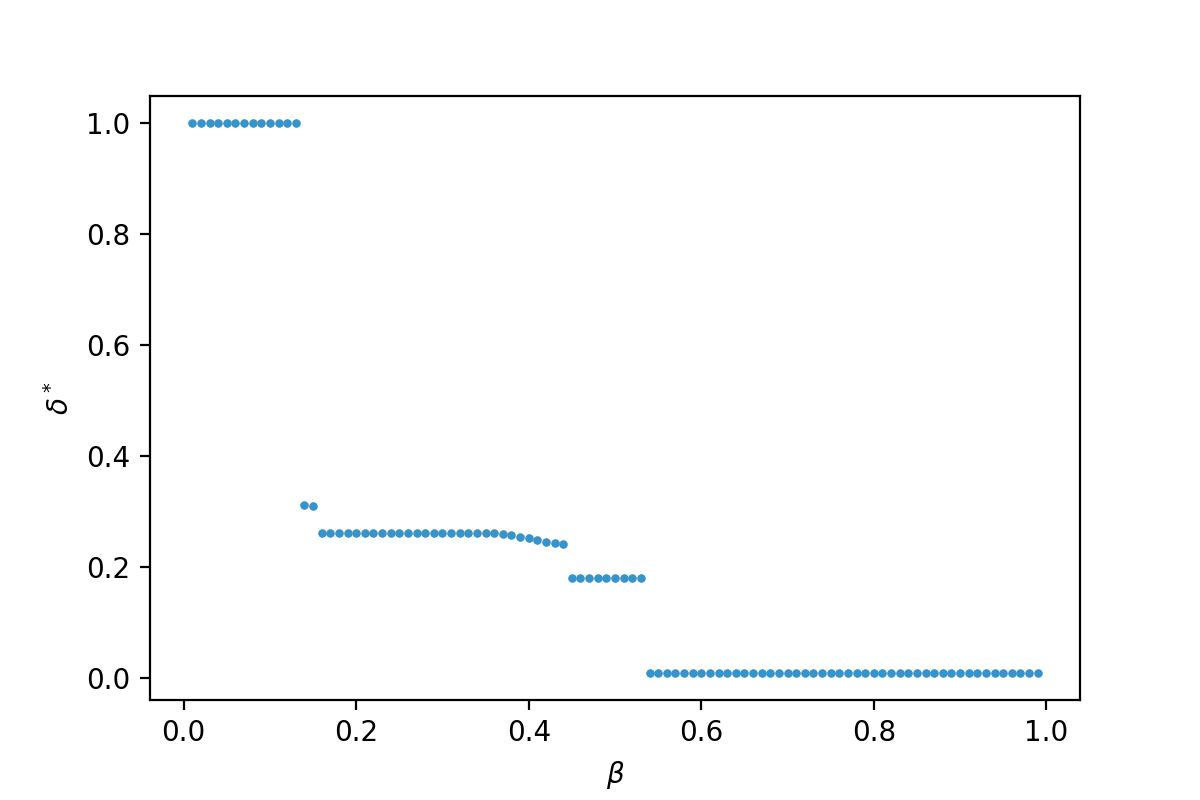} 
    \caption{The optimal $\delta^*$ with respect to $\beta$. Here, assume $X\sim\text{Exp}(\lambda)$. The insurer and the reinsurer adopt the power distortion risk measure with levels $\alpha$ and $\beta$, respectively.  Parameters are as follows. $\lambda=0.5$, $\theta_0=1$, $\theta_1=0.5$, $\theta_2=2$, $\alpha=0.2$.}
    \label{Fig:delta_beta_exp_power}
\end{figure}
\end{example}

\section{Conclusion}\label{sec:conclusion}

In this work, we introduce a reward-and-penalty variable premium scheme for reinsurance policies. Unlike classical premium principles, which assign a fixed value based on the distribution of a random loss, the proposed scheme not only accounts for the loss distribution, but also incorporates the realized loss amount. This feature introduces additional randomness for both the insurer and the reinsurer.
We begin by discussing the properties of this variable premium scheme and comparing them with the desirable characteristics of the classical premium principles. Next, we establish an optimization framework for the insurer and characterize the optimal reinsurance under mild assumptions regarding the insurer's risk measure.
In particular, when the insurer uses distortion risk measure or even TVaR to quantify her risk exposure, we can further formulate the optimal solutions. 
Moreover, we formulate a Bowley optimization problem to capture the interaction between the insurer and the reinsurer. Through various numerical examples, we demonstrate that the reinsurer can benefit from adopting the variable premium scheme, as it significantly reduces her total risk exposure compared to the expected-value premium principle, which serves as a limiting case of the proposed variable premium scheme. 

In a future study, it is interesting to incorporate  the return into optimization problems. For example, one can adopt the risk-adjusted liability to formulate the insurer's optimization problem and the Bowley reinsurance problem between an insurer and a reinsurer.

\appendix\label{A}

\section{Appendix}

\subsection{Proofs in Section \ref{sec:opt sol with general RM}}\label{app:proof in sec3.1}

\begin{proof} [\textbf{Proof of Lemma} \ref{lem:I_opt_3layer}]
    For $I\in\mathcal I$, if $\mathbb E[I(X)] = 0$, then $I(X) = 0$ almost surely. This indemnify function is an element in $\mathcal S_3$.
    When $\delta = 0$, we have $\theta_1 = \theta_0$. Then $\pi(I(x)) = (1+\theta)\mathbb E[I(X)]$ is the expected value principle. Consequently, the problem \eqref{insurer's prob with variable prem} is reduced to the problem \eqref{Proj2_prob1}. By Theorem 6.1 in \cite{van1989optimal}, the optimal solution to the problem \eqref{Proj2_prob1} belongs to the class of stop-loss functions $I_d(x) = (x-d)_+$, where $d\geq0$. Clearly, we have $I_d(x)\in \mathcal{S}_3$. 
   
    In the following part, we assume $\mathbb E[I(X)]>0$, $\delta>0$, $\theta_1 \leq \theta_0 < \theta_2$.  
    Denote $x_d^I = \sup\{x \geq 0 : I(x)\leq d_I \}$ and $x_u^I = \sup\{x \geq 0 : I(x)\leq u_I\}$. Due to the continuity of $I(x)$, it is clear that $0 \leq x_d^I < x_u^I$. Then the premium \eqref{premium_I} can be further written as
    \begin{equation}\label{premium_x}
    \Pi_{I}(x) = 
    \begin{cases}
        (1+\theta_1)\mathbb E[I(X)], & x \leq x_d^I,\\
        \delta I(x) + (1+\theta_0-\delta)\mathbb E[I(X)], & x_d^I < x \leq x_u^I,\\
        (1+\theta_2)\mathbb E[I(X)], &x > x_u^I.
    \end{cases}
    \end{equation}

    Arbitrarily select and fix $I\in\mathcal I$, we are going to show that there exists $f_I \in\mathcal S_3$ such that $\mathbb E[f_I(X)] = \mathbb E[I(X)]$ and $T_I$ is dominated by $T_{f_I}$ in the sense of convex order.
    
    \begin{enumerate}[(i)]
        \item If  $x_d^I < x_u^I < \infty$, define
        \begin{equation}\label{construct_f1}
            f_1(x) = \left\{
            \begin{aligned}
                &I(x), &x\leq x_u^I,\\
                &I(x_u^I) + (x-e)_+, &x>x_u^I,
            \end{aligned}
            \right.
        \end{equation}
        where $e\in[x_u^I, \infty)$ is determined by $\mathbb E[f_1(X)] = \mathbb E[I(X)]$. Such $e$ exists because we note that if $ e = x_u^I $, then $f_1 (x)\geq I(x)$ for all $x \geq 0$; if $ e = \infty $, then $ f_1 (x) \leq I(x) $ for all $x \geq 0$. Together with the continuity of $\E [f_1 (X) ]$ as a function of $e$, we can verify that there exists $ e $ such that $\mathbb E[f_1(X)] = \mathbb E[I(X)]$. It then follows that
        \begin{align}\label{Ef1_EI}
            \mathbb E[f_1(X)] 
            & = \mathbb E[f_1(X)|X\leq x_u^I]\mathbb P(X\leq x_u^I) + \mathbb E[f_1(X)|X > x_u^I]\mathbb P(X > x_u^I) \notag \\
            & = \mathbb E[I(X)|X\leq x_u^I]\mathbb P(X\leq x_u^I) + \mathbb E[f_1(X)|X > x_u^I]\mathbb P(X > x_u^I) \\
            & = \mathbb E[I(X)|X\leq x_u^I]\mathbb P(X\leq x_u^I) + \mathbb E[I(X)|X > x_u^I]\mathbb P(X > x_u^I) = \mathbb E[I(X)].\notag
        \end{align}
        Note that $x_u^I<\infty$ implies $\mathbb P(X>x_u^I)>0$, we obtain from \eqref{Ef1_EI} that 
        \begin{equation}\label{equation_f1}
            \mathbb E[f_1(X)|X > x_u^I] = \mathbb E[I(X)|X > x_u^I].
        \end{equation}
        In addition, $d_{f_1} = \frac{\theta_1 - \theta_0 + \delta}{\delta}\mathbb E[f_1(X)] = \frac{\theta_1 - \theta_0 + \delta}{\delta}\mathbb E[I(X)] = d_I$ and $u_{f_1} = \frac{\theta_2 - \theta_0 + \delta}{\delta}\mathbb E[f_1(X)] = \frac{\theta_2 - \theta_0 + \delta}{\delta}\mathbb E[I(X)] = u_I$, implying $x_d^{f_1} = x_d^I < \infty$ and $x_u^{f_1} = e \in[x_u^I, \infty)$. By \eqref{premium_x}, \eqref{construct_f1}, and \eqref{equation_f1}, we obtain
        \begin{align*}
            \mathbb E[\Pi_I(X)] 
            & = \mathbb E[\Pi_I(X)|X\leq x_u^I]\mathbb P(X\leq x_u^I) + \mathbb E[\Pi_I(X)|X > x_u^I]\mathbb P(X > x_u^I)\\
            & = \mathbb E[\Pi_{f_1}(X)|X\leq x_u^I]\mathbb P(X\leq x_u^I) + (1+\theta_2)\mathbb E[f_1(X)]\mathbb P(X > x_u^I)\\
            & = \mathbb E[\Pi_{f_1}(X)|X\leq x_u^{f_1}]\mathbb P(X\leq x_u^{f_1}) + (1+\theta_2)\mathbb E[f_1(X)]\mathbb P(X > x_u^{f_1}) = \mathbb E[\Pi_{f_1}(X)].
        \end{align*}
        The third equation holds because $\Pi_{f_1}(x) = (1+\theta_2)\mathbb E[f_1(X)]$ for $x\in[x_u^I, x_u^{f_1}]$. It then follows that $\mathbb E[T_I(X)] = \mathbb E[R_I(X) + \Pi_I(X)] = \mathbb E[R_{f_1}(X) + \Pi_{f_1}(X)] = \mathbb E[T_{f_1}(X)]$. Since $1-\delta\geq0$, we can find that $T_I(x)$ up-crosses $T_{f_1}(x)$. By Lemma 3 in \cite{ohlin1969class}, we have
        \begin{equation}\label{cx1}
            T_{f_1}(X)\leq_{cx}T_I(X).
        \end{equation}

        Then, we define
        \begin{equation*}
            f_2(x) = \left\{
            \begin{aligned}
                &f_1(x_d^{f_1}) + (x-c)_+ - (x-(c + u_{f_1} - d_{f_1}))_+, &x_d^{f_1}\leq x < x_u^{f_1},\\
                &f_1(x), &\text{otherwise},
            \end{aligned}
            \right.
        \end{equation*}        
        where $c\in[x_d^{f_1}, x_u^{f_1})$ is determined by $\mathbb E[f_2(X)] = \mathbb E[f_1(X)]$. By similar argument, we obtain
        \begin{equation}\label{cx2}
            T_{f_2}(X)\leq_{cx}T_{f_1}(X).
        \end{equation}
        
        Finally, define
        \begin{equation*}
            f_3(x) = \left\{
            \begin{aligned}
                &(x-a)_+\wedge d_{f_2}, &x\leq x_d^{f_2},\\
                &f_2(x), & x > x_d^{f_2},
            \end{aligned}
            \right.
        \end{equation*}          
        where $a\in[0,x_d^{f_2})$ is defined by $\mathbb E[f_3(X)] = \mathbb E[f_2(X)]$. By similar argument, we have
        \begin{equation}\label{cx3}
            T_{f_3}(X)\leq_{cx}T_{f_2}(X).
        \end{equation}
        Combining \eqref{cx1}, \eqref{cx2}, and \eqref{cx3}, it then follows that $T_{f_3}(X)\leq_{cx}T_{I}(X)$. 
        
        In this case, based on given $I$, we can always construct $f_I\in\mathcal S_3$  in the form of 
        \begin{equation*}\label{f_3layer}
            f_I =  (x-a)_+ - (x-b)_+ + (x-c)_+ - (x-d)_+ + (x-e)_+
        \end{equation*}
        with parameters $0\leq a\leq b = a+d_I\leq c\leq d=c+u_I- d_I\leq e$ such that $\mathbb E[f_I(X)] = \mathbb E[I(X)]$ and $T_{f_I}(X)\leq_{cx}T_I(X)$.

        \item If $x_d^I < x_u^I = \infty$, following the same argument, we can find $f_I\in\mathcal S_3$ in the form of $$f_I = (x-a)_+ - (x-b)_+ + (x-c)_+ - (x-d)_+$$
        with parameters $0\leq a\leq b = a + d_I\leq c\leq d = c + u_I - d_I$, such that $\mathbb E[f_I(X)] = \mathbb E[I(X)]$ and $T_{f_I}(X)\leq_{cx}T_I(X)$.

        \item If $x_d^I = \infty$, we have $I(x)\leq d_I = \frac{\theta_1 - \theta_0 + \delta}{\delta}\mathbb E[I(X)] \leq\mathbb E[I(X)]$. This means $I(x) = 0$ almost surely, which is an element in $\mathcal S_3$.
    \end{enumerate}
    Combining the three cases, we conclude that for any $I\in\mathcal I$, there always exists $f_I\in\mathcal S_3$ such that $\mathbb E[f(X)] = \mathbb E[I(X)]$ and $T_{f_I}(X)\leq_{cx}T_I(X)$. The convex order consistency of $\rho$ gives $\rho(T_{f_I}(X))\leq \rho(T_I(X))$, which completes the proof.
\end{proof}

\begin{proof}[\textbf{Proof of Theorem} \ref{Proj2_thm1}]
Using the similar argument in the proof of Lemma \ref{lem:I_opt_3layer}, we only need to consider non-trivial cases when $\mathbb E[I(X)]>0$, $\delta>0$, and $\theta_1 \leq \theta_0 < \theta_2$.

     For any $I\in\mathcal S_3$, we are going to show that there exists a function $h\in\tilde{\mathcal I}$ satisfying $\mathbb E[h(X)] = \mathbb E[I(X)]$ and $T_h(X)\leq_{cx}T_I(X)$.
    Define
    \begin{equation}\label{2layer_h1}
        h_1(x) = \left\{
        \begin{aligned}
            &I(x), & x\leq x_d^I,\\
            &I(x_d^I) + (x - c_1)_+, & x>x_d^I,
        \end{aligned}
        \right.
    \end{equation}
    where $c_1$ is defined by $\mathbb E[h_1(X) ] = \mathbb E[I(X)]$. It is easy to show $c_1\in[x_d^I, x_u^I-u_I+d_I]$. In specific, if $c_1 = x_d^I$, we have $h_1(x) \geq I(x)$ and thus $\mathbb E[h_1(X)] \geq \mathbb E[I(X)]$. If $c_1 = x_u^I-u_I+d_I$, we have $h_1(x) \leq I(x)$ and thus $\mathbb E[h_1(X)] \leq \mathbb E[I(X)]$. Since $\mathbb E[h_1(X)]$ is a continuous function of $c_1$, there must exists $c_1\in[x_d^I, x_u^I-u_I+d_I]$ such that $\mathbb E[h_1(X)] = \mathbb E[I(X)]$. 
    Therefore, we have $d_{h_1} = \frac{\theta_1-\theta_0+\delta}{\delta}\mathbb E[h_1(X)] = d_I$ and $u_{h_1} = \frac{\theta_2-\theta_0+\delta}{\delta}\mathbb E[h_1(X)] = u_I$. In addition, we have $x_d^I\leq x_d^{h_1} = c_1 \leq x_u^{h_1} = c_1 + u_I-d_I \leq x_u^I.$
    For $f\in\{ I, h_1 \}$, by \eqref{premium_x}, we have
    \begin{equation*}
    T_f(x) = R_f(x) + \Pi_{f}(x) = 
    \begin{cases}
        x - f(x) + (1+\theta_1)\mathbb E[f(X)], & x \leq x_d^f,\\
        x - (1 - \delta) f(x) + (1+\theta_0-\delta)\mathbb E[f(X)], & x_d^f < x \leq x_u^f,\\
        x - f(x) + (1+\theta_2)\mathbb E[f(X)], &x < x_u^f.
    \end{cases}
    \end{equation*}
    It then follows that
    \begin{equation}\label{ET_ET}
    \begin{aligned}
        \mathbb E[T_I(X)] - \mathbb E[T_{h_1}(X)] 
        = &\, \mathbb E[h_1(X)-I(X)|X\leq x_d^I]\mathbb P(X\leq x_d^I)\\
        &\, + (1-\delta)\mathbb E[h_1(X)-I(X)|x_d^I < X < x_u^{h_1}]\mathbb P(x_d^I < X < x_u^{h_1})\\
        & \,+ \mathbb E[h_1(X)-I(X)|X \geq x_u^{h_1}]\mathbb P(X \geq x_u^{h_1}) \\
        \geq & \,\mathbb E[h_1(X)-I(X)] = 0.
    \end{aligned}
    \end{equation}
    The inequality holds because $h_1(x)\leq I(x)$ when $x\in[x_d^I, x_u^{h_1}]$. Thus, we have 
    \begin{equation}\label{EI>Eh}
        \mathbb E[T_I(X)] \geq \mathbb E[T_{h_1}(X)],
    \end{equation}
    where the equality holds if and only if $h_1(x) = I(x)$ for $x\geq0$. In addition, it is easy to find $T_I(X)$ up-crosses $T_{h_1}(X)$.
    
    We then define
    \begin{equation}\label{2layer_h2}
        h_2(x) = \left\{
        \begin{aligned}
            &(x-c_2)_+ - (x-c_2-u_I)_+, &x \leq x_u^I\\
            &I(x), &x > x_u^I.
        \end{aligned}
        \right.
    \end{equation}
    Here, $c_2$ is determined by $\mathbb E[h_2(X)] = \mathbb E[I(X)]$. If $c_2 = 0$, we have $h_2(x)\geq I(x)$ and thus $\mathbb E[h_2(X)]\geq \mathbb E[I(X)]$. If $c_2 = x_d^I-d_I$, we have $h_2(x)\leq I(x)$ and thus $\mathbb E[h_2(X)]\leq \mathbb E[I(X)]$. Because $\mathbb E[h_2(X)]$ is continuous in $c_2$, there must exists $c_2$ such that $\mathbb E[h_2(X)] = \mathbb E[I(X)]$. Therefore, $c_2\in[0,x_d^I-d_I]$. 

    Then, we have $d_{h_2} = \frac{\theta_1-\theta_0+\delta}{\delta}\mathbb E[h_2(X)] = d_I$ and $u_{h_2} = \frac{\theta_2-\theta_0+\delta}{\delta}\mathbb E[h_2(X)] = u_I$. It then follows that $x_d^{h_2} = c_2 + d_{h_2}\leq x_d^I\leq x_u^I$. Thus, we have
    \begin{equation}
    \begin{aligned}
        \mathbb E[T_I(X)] - \mathbb E[T_{h_2}(X)] 
        = & \,\mathbb E[h_2(X)-I(X)|X\leq x_d^{h_2}]\mathbb P(X\leq x_d^{h_2})\\
        & \,+ (1-\delta)\mathbb E[h_2(X)-I(X)|x_d^{h_2} < X < x_u^I]\mathbb P(x_d^{h_2} < X < x_u^I)\\
        &\, + \mathbb E[h_2(X)-I(X)|X \geq x_u^I]\mathbb P(X \geq x_u^I) \\
        \leq &\, \mathbb E[h_2(X)-I(X)] = 0.
    \end{aligned}
    \end{equation}
    The inequality holds because $h_2(x)\geq I(x)$ in $[x_d^{h_2}, x_u^I]$. It then follows that
    \begin{equation}\label{EI<Eh}
        \mathbb E[T_I(X)] \leq \mathbb E[T_{h_2}(X)].
    \end{equation}
    Here, the equality holds if and only if $h_2(x) = I(x)$ for $x\geq0$. In addition, $T_I(X)$ up-crosses $T_{h_1}(X)$.
    
    Combining \eqref{EI>Eh} and \eqref{EI<Eh}, we can conclude that, for an arbitrarily selected $I\in\mathcal S_3$, if $I\in\tilde{\mathcal I}$, then \eqref{I_opt_2layer} holds and completes the proof. If $I\in S_3\cap\tilde{\mathcal I}^c$, then there exists $h_1\in\mathcal I_1$ and $h_2\in\mathcal I_2$ such that $\mathbb E[h_1(X)] = \mathbb E[h_2(X)] = \mathbb E[I(X)]$ and $\mathbb E[T_{h_1}(X)]<\mathbb E[T_I(X)]<\mathbb E[T_{h_2}(X)]$. In the next part, we will show that, based on $h_1$ and $h_2$, there exists $h\in\tilde{\mathcal I}$ such that $\mathbb E[h(X)] = \mathbb E[I(X)]$ and $\mathbb E[T_{h}(X)] = \mathbb E[T_I(X)]$.

    According to \eqref{2layer_h1}, we write $h_1(x) = (x-x_1)_+ - (x-x_2)_+ + (x-x_3)_+$ with $0\leq x_1 < x_2 = x_1+d_{h_1} \leq x_3$. Define $\tilde h_1(x) = (x-\tilde x_1)_+ - (x-\tilde x_2)_+ + (x-\tilde x_3)_+$ with $x_1 < \tilde x_1 < \tilde x_2 = \tilde x_1 + d_{\tilde h_1}\leq \tilde x_3$ such that $\mathbb E[\tilde h_1(X)] = \mathbb E[h_1(X)]$. Here, $\tilde h_2\in\mathcal I_1$. Then, we have $d_{\tilde h_1} = \frac{\theta_1 - \theta_0 + \delta}{\delta}\mathbb E[\tilde h_1(X)] = \frac{\theta_1 - \theta_0 + \delta}{\delta}\mathbb E[h_1(X)] = d_{h_1}$. For given $x_1<\tilde x_1$, it is easy to find that $\tilde x_3 < x_3$. By the similar argument, we can prove that $\mathbb E[T_{\tilde h_1}(X)] > \mathbb E[T_{h_1}(X)]$ and $T_{h_1}$ up-crosses $T_{\tilde h_1}$. With given $\mathbb E[\tilde h_1(X)]$, $\tilde x_2$ and $\tilde x_3$ can be determined by $\tilde x_1$, and thus $\mathbb E[T_{\tilde h_1}(X)]$ can be viewed as an increasing function of $\tilde x_1$. 
    
    By \eqref{2layer_h2}, write $h_2(x) = (x-y_1)_+ - (x-y_2)_+ + (x-y_3)_+$ with $0\leq y_1 < y_2 = y_1+u_{h_2} \leq y_3$. Define $\tilde h_2(x) = (x-\tilde y_1)_+ - (x-\tilde y_2)_+ + (x-\tilde y_3)_+$ with $y_1 < \tilde y_1 < \tilde y_2 = \tilde y_1 + u_{\tilde h_2}\leq \tilde y_3$ such that $\mathbb E[\tilde h_2(X)] = \mathbb E[h_2(X)]$. Here, $\tilde h_2\in\mathcal I_2$. Then, we have $u_{\tilde h_2} = \frac{\theta_2 - \theta_0 + \delta}{\delta}\mathbb E[\tilde h_2(X)] = \frac{\theta_2 - \theta_0 + \delta}{\delta}\mathbb E[h_2(X)] = u_{h_2}$. For given $y_1 < \tilde y_1$, we can find that $\tilde y_3 < y_3$. Similarly, we have $\mathbb E[T_{\tilde h_2}(X)] < \mathbb E[T_{h_2}(X)]$ and $T_{h_2}$ up-crosses $T_{\tilde h_2}$. Since $\tilde y_2$ and $\tilde y_3$ can be determined by $\tilde y_1$ under given $\mathbb E[\tilde h_2(X)]$, we then conclude that $\mathbb E[T_{\tilde h_2}(X)]$ is a decreasing function of $\tilde y_1$. 
    
    Define a stop-loss function $I_d(x) = (x-d)_+$ such that $\mathbb E[I_d(X)] = \mathbb E[I(X)]$. Note that $I_d$ belongs to both $\mathcal I_1$ and $\mathcal I_2$, and 
    \begin{equation}
        \mathbb E[T_{I_d}(X)] = \max_{\tilde h_1\in\mathcal I_1, ~\mathbb E[\tilde h_1(X)] = \mathbb E[I(X)] }\mathbb E[T_{\tilde h_1}(X)] = \min_{\tilde h_2\in\mathcal I_2, ~\mathbb E[\tilde h_2(X)] = \mathbb E[I(X)]}\mathbb E[T_{\tilde h_2}(X)].
    \end{equation}
    This means for $\mathbb E[T_I(X)]\in(\mathbb E[T_{h_1}(X)], \mathbb E[T_{h_2}(X)])$, there exists $f_I(x)\in\tilde{\mathcal I} = \mathcal I_1\cup\mathcal I_2$ such that $\mathbb E[f_I(X)] = \mathbb E[I(X)]$ and $\mathbb E[T_{f_I}(X)] = \mathbb E[T_I(X)]$. Furthermore, since $T_I(X)$ up-crosses $T_{\tilde h_1}(X)$ and $T_{\tilde h_2}(X)$ for any $\tilde h_1\in\mathcal I_1$ and $\tilde h_2\in\mathcal I_2$, by Lemma 3 in \cite{ohlin1969class}, we have $T_I(X)\geq_{cx}T_{f_I}(X)$, which completes the proof.
\end{proof}

\bibliographystyle{elsarticle-harv}
\bibliography{mybib.bib}
\end{document}